\documentclass[amsmath,10pt,twocolumn,superscriptaddress,footinbib,pre]{revtex4-1}
\usepackage{amsmath,amsfonts,bm}
\usepackage{graphicx}
\usepackage{color}
\usepackage{subfigure}

\begin{document}
\title{Heterogeneity-induced large deviations in activity and (in some cases) entropy production}
\author{Todd R. Gingrich}
\affiliation{Department of Chemistry, University of California, Berkeley, 
California 94720}
\author{Suriyanarayanan Vaikuntanathan}
\affiliation{Material Sciences Division, Lawrence Berkeley National 
Laboratory, Berkeley, California 94720}
\author{Phillip L. Geissler}
\email[]{geissler@berkeley.edu}
\affiliation{Department of Chemistry, University of California, Berkeley, 
California 94720}
\affiliation{Material Sciences Division, Lawrence Berkeley National 
Laboratory, Berkeley, California 94720}
\affiliation{Chemical Sciences Division, Lawrence Berkeley National 
Laboratory, Berkeley, California 94720}

\begin{abstract}

We solve a simple model that supports a dynamic phase transition and show conditions for the existence of the transition.
Using methods of large deviation theory we analytically compute the probability distribution for activity and entropy production rates of the trajectories on
a large ring with a single heterogeneous link.
The corresponding joint rate function demonstrates two dynamical phases \textemdash \ one localized and the other delocalized, but the marginal rate functions do not always exhibit the underlying transition.
Symmetries in dynamic order parameters influence the observation of a transition, such that distributions for certain dynamic order parameters need not reveal an underlying dynamical bistability. 
Solution of our model system furthermore yields the form of the effective Markov transition matrices that generate dynamics in which the two dynamical phases are at coexistence.
We discuss the implications of the transition for the response of bacterial cells to antibiotic treatment, arguing that even simple models of a cell cycle lacking an explicit bistability in configuration space will exhibit a bistability of dynamical phases.

\end{abstract}
\maketitle 

\section{Introduction}

Large dynamical fluctuations and dynamical heterogeneity are characteristic features of non-equilibrium chemical and biological systems.
Important examples include fluctuations in currents and transport properties of molecular motors~\cite{Depken2013}, dynamic instability in actin and microtubule growth~\cite{Brun2009,Dogterom1993}, dynamical heterogeneity in cell migration~\cite{Angelini2011}, and intermittency in cell growth rates~\cite{Balaban2004}.
Large deviation theories and statistical mechanics on the level of trajectories provide convenient frameworks to characterize the dynamical fluctuations.
Of particular interest to the present work are the emergence of dynamic phases analogous to the emergence of phases in the conventional statistical mechanics of first order phase transitions~\cite{Touchette2009, Garrahan2009}.
The existence of dynamic phases indicates that the most probable trajectories naturally cluster into classes with distinct dynamical properties.
As in equilibrium statistical mechanics, a very productive perspective on the origins and consequences of dynamic phase transitions can be gained by scrutinizing the statistics of pertinent order parameters.
A hump or ``fat tail'' in the wings of such distributions can reveal the presence of a second dynamical phase.
Importantly, the separation of trajectories into distinct classes can be made rigorous by demonstrating a singularity in the appropriate scaled cumulant generating function~\cite{Touchette2009}.
Conversely, demonstrating such a singularity implies a broad distribution for dynamical fluctuations, which can rationalize experimentally observed dynamical heterogeneity.

This perspective has been elaborated in several interesting contexts~\cite{Espigares2013,Bodineau2004}, albeit in most cases for complicated many-body systems that do not permit full analytical solutions and are thus not entirely transparent.
Notable examples include lattice and molecular models of glasses~\cite{Bodineau2012,Hedges2009}, asymmetric exclusion processes~\cite{Derrida2007} and zero-range processes~\cite{Harris2005}.
It has recently been shown that similarly complex behavior can emerge in seemingly very simple systems which do permit an analytical treatment.
Specifically, we have shown that a dynamic phase transition can be demonstrated analytically for a biased random walker on a ring with a single impurity in the transition rates~\cite{Vaikuntanathan2014}. 
The relatively simple analytics of our random walker model provides an excellent arena for addressing two basic questions about dynamic phase transitions.
Firstly, under what conditions will a dynamic phase transition emerge?
Secondly, are there physical methods for modulating the dynamics to achieve coexistence between dynamical phases or to induce transitions between them?

To this end, here we investigate the statistics of two dynamical order parameters whose fluctuations can reveal the phase transition.
We analytically construct the joint rate function for entropy production (Eq.~\ref{eq:entropyproductiondef}) and dynamical activity~\cite{Garrahan2007} (Eq.~\ref{eq:activitydef}), which is analogous to a two dimensional free energy surface.
The two-dimensional rate function in all cases reveals two basins, corresponding to two distinct classes of trajectories, one localized and the other delocalized.
However, when one of the order parameters is integrated out, the remaining marginal distribution does not necessarily reveal the underlying bistability.
In particular, we show regimes for which the dynamical activity statistics is influenced by two dynamical phases while the fluctuations in entropy production reveal only a single phase.

The dynamic phase transition implies the existence of a rare localized class of trajectories~\cite{Vaikuntanathan2014}.
We investigate conditions required to induce the transition, thereby causing the localized trajectories to become typical.
In conventional statistical mechanics a rare phase can be made dominant by adjusting intensive fields like temperature, pressure, or chemical potential.
The statistical mechanics of trajectories is more complicated as the field conjugate to a dynamical order parameter (the $\lambda$ or $s$ field throughout this paper) is time-non-local and therefore cannot be experimentally tuned in a straightforward way.
We construct Markov matrices which (in the long time limit) are equivalent to the natural dynamics with a $\lambda$ or $s$ field~\cite{Jack2010,Chetrite2013}.
These Markov matrices reveal the \emph{physical} values of the rate constants which would place the two dynamical phases at coexistence.
In other words, with the computed set of rate constants, long trajectories switch back and forth equally between localized and delocalized behavior. 
Using the Markov matrices that generate effective $\lambda$ field dynamics, we also show that $\lambda$ field biasing cannot induce non-equilibrium currents that violate detailed balance.
These biasing techniques can amplify (or suppress) existing non-equilibrium currents, but when applied to an equilibrium system the methods simply transform from one detail balanced dynamics to another.

Finally, we consider the problem of observed heterogeneity in cell growth rates and apply results from our model system to elucidate this phenomenon. 
In particular, it has been observed that a stochastic subpopulation of cells in an \textit{E. Coli} colony exhibit markedly reduced growth rates~\cite{Balaban2004}. 
These cells, labeled persisters, are more likely to survive antibiotic treatment~\cite{Lewis2010}. 
Treating our model system as an extremely simplified version of the cell growth cycle, we argue that the phenomenon of persistence should be a generic consequence of a class of localized trajectories that is rare in the absence of antibiotics. 
We show how treating cells with different strengths of antibiotics in experiments might be equivalent to effectively tuning a $\lambda$ field and induce a transition between different dynamical behaviors. 
We also note that coexistence in the space of trajectories can facilitate massive dynamical fluctuations which are evocative of those observed in other biological contexts such as growing polymers including microtubules~\cite{Mitchison1984}, actin~\cite{Fujiwara2002}, and bacterial homologs thereof~\cite{Garner2004, Popp2010, Dimitrov2008}, where trajectories exhibit a stark switching between growing and collapsing behaviors.  
Our work clarifies the conditions required for such phase coexistence in trajectories and also illuminates the properties of the phase transition. 

The structure of the paper is as follows.
In Section \ref{sec:framework} we review the basic structure of the large deviation calculations.
We then introduce our solvable model system in Section \ref{sec:scgf} and derive the scaled cumulant generating function, the Legendre transform of which yields the entropy production and activity statistics.
Using this result, we discuss in Section \ref{sec:props} the nature of a dynamic phase transition and the conditions for which the transition can be observed by these order parameters.
Finally, we address implications of such a dynamic phase transition.
We both identify conditions for dynamical coexistence in which the two phases contribute equally and discuss the way in which the response of cells to antibiotic treatment may expose a similar underlying transition.

\section{Framework}
\label{sec:framework}
We consider continuous-time Markovian dynamics on a discrete state space.
Such a stochastic dynamics is compactly represented by a master equation with rate matrix $\mathbb{W}$ whose off-diagonal elements $\mathbb{W}_{ij}$ detail the rates of transition from state $j$ to state $i$~\cite{vankampen1992}.
The probability distribution of the set of all possible trajectories is well-defined in the steady-state.
We investigate both typical and rare dynamical fluctuations by considering the behavior of dynamic order parameters.

Time-additive dynamic order parameters are particularly relevant to many experiments as they report on cumulative dynamical behavior, for example the net current observed in a finite time experiment.
In this paper we consider two such order parameters, the entropy production and the dynamical activity.
The entropy production of a trajectory is defined in the stochastic thermodynamics sense as the log ratio of forward and reverse probabilities~\cite{Seifert2012}.
We focus on continuous time hopping processes, in which case the entropy production can be expressed as
\begin{equation}
\omega = \sum_{\text{hops}} \ln \frac{k_{\text{f}}}{k_{\text{r}}},
\label{eq:entropyproductiondef}
\end{equation}
where $k_\text{f}$ and $k_\text{r}$ are the forward and reverse rate constants for each hop.
The dynamical activity, $K$, simply counts the total number of hops.
\begin{equation}
K = \sum_{\rm hops} 1
\label{eq:activitydef}
\end{equation}
This accounting of microscopic transitions has been used most predominantly in the study of glassy dynamics~\cite{Garrahan2007}.
Whereas the entropy production provides a measure of the dissipation associated with a trajectory, the dynamical activity indicates how labile the dynamics is.
By considering both order parameters we highlight how the statistics of various observables may be differently affected by the dynamic phase transition.

\begin{figure*}[tbhp]         
\centering
\includegraphics[width=0.98\linewidth]{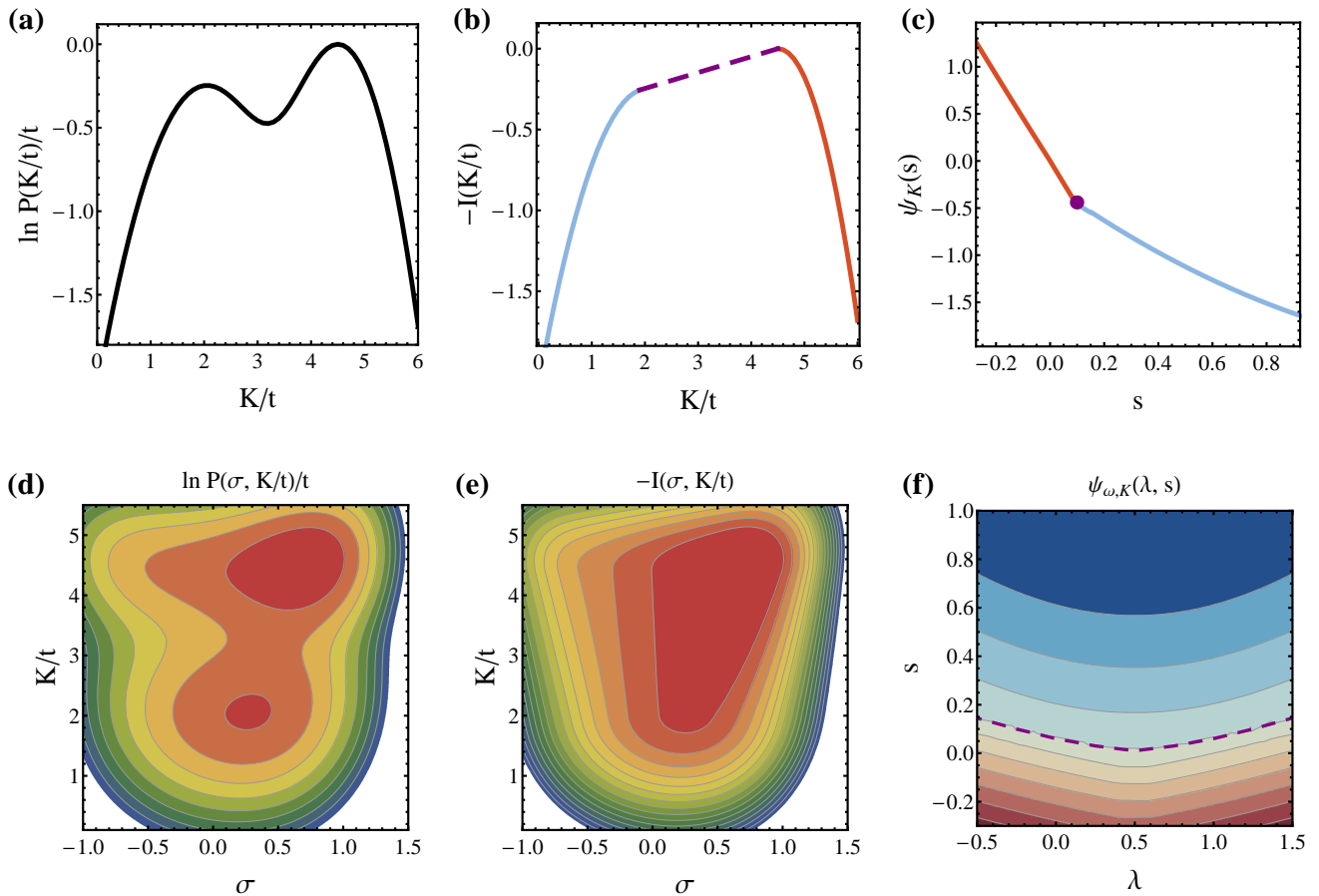}
\caption{(Color online) Schematic illustrating the relationships between order parameter probability distributions collected in a finite time experiment (a, d), rate functions (b, e), and cumulant generating functions (c, f) in one and two dimensions.  
The rate function is the concave (for our sign convention) hull of the finite time distribution and the Legendre transform of the scaled cumulant generating function. 
The bimodality of the finite time distribution results in a so-called tie line in the concave hull, represented as a dashed line in the 1d rate function.
The tie line necessitates a singularity in the scaled cumulant generating function, depicted as a dot in (c) and as a dashed line in (f).
Note that the 1d cumulant generating functions for activity (entropy production) are given by the $\lambda = 0 \ (s = 0)$ slice in (f).
}
\label{fig:cartoon}
\end{figure*} 

The time-additivity of these order parameters allows their probability distribution to be described by a large deviation form,
\begin{equation}
P(\sigma, K/t) \approx e^{-t I(\sigma, K/t)},
\label{eq:largedeviationprinciple}
\end{equation}
where $I(\sigma, K/t)$ is the joint large deviation rate function and $\sigma = \omega / t$ is the entropy production rate.
The rate function $I(\sigma, K/t)$ can be computed as the Legendre transform of the scaled cumulant generating function~\cite{Touchette2009, Lebowitz1999},
\begin{equation}
\psi_{\omega, K}(\lambda, s) = \lim_{t \rightarrow \infty} \frac{1}{t} \ln \left<e^{-\lambda \omega - s K}\right>,
\label{eq:cgf}
\end{equation}
where the expectation value is taken over trajectories initialized in the steady state distribution.
This function can in turn be obtained as the maximum eigenvalue of a tilted operator, $\mathbb{W}_{\omega, K}(\lambda, s)$, which is simply related to $\mathbb{W}$~\cite{Lebowitz1999}.
Specifically the matrix elements are given by
\begin{equation}
\mathbb{W}_{\omega, K}(\lambda, s)_{ij} = \left(1 - \delta_{ij}\right)\mathbb{W}_{ij}^{1-\lambda} \mathbb{W}_{ji}^\lambda e^{-s} + \delta_{ij} \mathbb{W}_{ij}.
\label{eq:Woperatordef}
\end{equation}
By solving for the eigenspectrum of $\mathbb{W}_{\omega, K}(\lambda, s)$ we can thus compute the long time limit of $P(\sigma, K/t)$ via a Legendre transform.

Fig.~\ref{fig:cartoon} graphically illustrates the relationships between probability distributions for the dynamic order parameters, rate functions, and scaled cumulant generating functions.
In particular, the singularities in $\psi_{\omega, K}(\lambda, s)$ generate bistable order parameter distributions.
For the type of ergodic dynamics studied here these bistable distributions tend toward a strictly concave rate function in the long time limit~\cite{Touchette2013}, which is the Legendre transform of the scaled cumulant generating function with a Maxwell construction.
Despite the underlying bistability of the distribution shown in the Fig.~\ref{fig:cartoon}(d), the marginals of that distribution need not illustrate a bistability, if the two basins are appropriately aligned.
We now shift our attention to a particular solvable model, whereby computing $\psi_{\omega, K}(\lambda, s)$ we can determine conditions for phase transitions in the two order parameters.

\section{Analytic Solution to 1d Random Walker on a Ring with a Heterogeneity}
\label{sec:scgf}

\begin{figure}[tbhp]         
\centering
\includegraphics[width=0.98\linewidth]{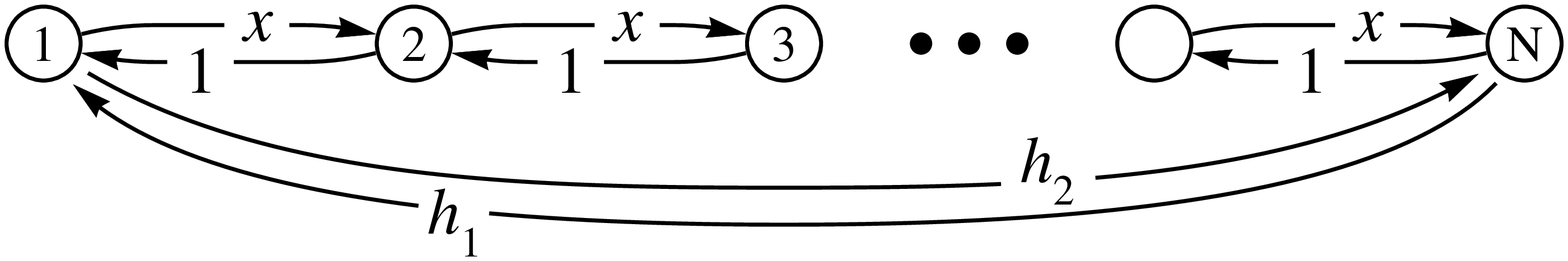}
\caption{Network of states and the rates for transitioning between the states.  For certain choices of $x$, $h_1$, and $h_2$ we demonstrate a dynamic phase transition in entropy production and dynamical activity rates.}
\label{fig:ringnetwork}
\end{figure}

We consider dynamics of a single particle on a network of $N$ states arranged in a ring as depicted in Fig.~\ref{fig:ringnetwork}.
Clockwise rates are given by $x$ and counterclockwise rates by $1$ except for the rates at a single heterogeneous link, which are given by $h_1$ and $h_2$, respectively.
A trajectory on the network corresponds to a sequence of hops from state to state with a Poisson-distributed waiting time between hops determined by the rate constants $x$, $h_1$, and $h_2$.
Without loss of generality we focus on the case that $x > 1$ such that typical trajectories cycle in the clockwise direction.
For generic choices of $h_1$ and $h_2$ the dynamics is out-of-equilibrium, which can be seen most simply since the probability of clockwise cycles differs from that of counter-clockwise cycles~\cite{Schnakenberg1976}.
Because the network supports only a single cycle, it is one of the simplest models for non-equilibrium dynamics.
This simplicity enables an analytic solution in the limit that the ring grows infinitely large.
Because analytically solvable non-equilibrium models are few, these solutions can provide a useful reference point.

We focus on a large $N$ limit which maintains the discrete nature of the states.
This limit is appropriate for chemical reaction kinetics with transitions among a discrete set of states~\cite{vankampen1992}.
It may also be interesting to consider continuum limits with rates scaled by $N$.
Such a continuum limit is pertinent to a Brownian particle confined to a ring~\cite{Gomez2009}, and has been discussed in a Freidlin-Wentzell framework without a heterogeneity in the limit of small noise~\cite{Faggionato2012}.

Following the framework of Section~\ref{sec:framework}, the tilted operator can be written down straightforwardly as
\begin{equation}
\mathbb{W}_{\omega, K}(\lambda, s) = 
\begin{pmatrix}                                                                
-x - h_2 & x^{\lambda}e^{-s} & \hdots & h_1^{1-\lambda}h_2^{\lambda}e^{-s}\\
x^{1-\lambda}e^{-s} & -1-x & \hdots & 0\\
\vdots & \vdots & \ddots & \vdots\\
h_1^{\lambda}h_2^{1-\lambda}e^{-s} & 0 & \hdots & -h_1-1
\end{pmatrix}.
\label{eq:tiltedop}
\end{equation}
For modest $N$ one can numerically calculate the largest eigenvalue of this matrix to yield the scaled cumulant generating function $\psi_{\omega, K}(\lambda,s)$.
In the large $N$ limit, however, we can obtain an analytic form for the limiting behavior using a perturbation theory we recently outlined~\cite{Vaikuntanathan2014}.

Were it not for the heterogeneous link, there would be a translational symmetry allowing the tilted operator to be exactly diagonalized via a Fourier transform.
For $\lambda$ and $s$ in a particular region of the $(\lambda, s)$ plane, the maximum eigenvalue of the tilted operator in Eq.~\eqref{eq:tiltedop} coincides with this solution for the translationally symmetric network in the large $N$ limit.
$\psi_{\omega, K}(\lambda, s)$ exhibits a cusp along the boundary of this region.
One side of the boundary corresponds to a maximal right eigenvector of the tilted operator which is delocalized while on the other side the eigenvector is exponentially localized around the heterogeneous link.
The discontinuity of the slopes of $\psi_{\omega, K}(\lambda, s)$ when crossing this boundary indicates a dynamic phase transition between classes of trajectories which are localized and those which are delocalized.
The detailed calculation, provided in Appendix~\ref{app:jointcalc}, reveals that the line of cusps separating localized from delocalized eigenvectors is given by the logarithm of the roots of a quadratic,
\begin{widetext}
\begin{equation}
s^*(\lambda) = \ln \left(\frac{1 + x - h_1 - h_2 - \sqrt{(h_1 - x - h_2 + 1)^2 + 4(h_2-1)(h_1-x)h_1 h_2 x^{-2\left(\left|\lambda - \frac{1}{2}\right|+\frac{1}{2}\right)}}}{2(h_2-1)(h_1-x) x^{-\left(\left|\lambda - \frac{1}{2}\right| + \frac{1}{2}\right)}}\right)
\label{eq:sstar}
\end{equation}
\end{widetext}
This equation, corresponding to the purple dashed curve in the schematic of Fig.~\ref{fig:cartoon}(f), is plotted for particular choices of $x, h_1,$ and $h_2$ in Fig.~\ref{fig:psih1h2xvals}(a).

Remarkably, the value of $\psi_{\omega, K}(\lambda, s)$ everywhere can be determined by the solution to the translationally symmetric network and the form of $s^*(\lambda)$.
This follows since $\psi_{\omega, K}(\lambda, s)$ is continuous and the partial derivatives with respect to $\lambda$ must vanish in the localized regime~\footnote{Localized eigenvectors of the tilted operator correspond to localized trajectories, which cannot produce entropy in the long-time limit.  Partial derivatives with respect to $\lambda$ must then yield a zero entropy production.}.
The translationally symmetric network solution evaluated along the line of cusps thus provides the maximum eigenvalue in the localized region giving
\begin{equation}
\psi_{\omega, K}(\lambda, s) = \begin{cases}
x^{1-\lambda} e^{-s} + x^\lambda e^{-s} - 1 - x, & s \leq s^*\\
x^{1-\lambda^*} e^{-s} + x^{\lambda^*} e^{-s} - 1 - x, & s > s^*,
\end{cases}
\label{eq:jointcgf}
\end{equation}
where $s^*$ and $\lambda^*$ are shorthand for $s^*(\lambda)$ given in Eq.~\eqref{eq:sstar} and for the inverse function $\lambda^*(s)$~\footnote{There is not a unique inverse, but the sum $x^{1-\lambda^*} + x^{\lambda^*}$ is the same for either choice of the inverse.}.
This is our primary analytical result, which enables the computation of the probability distributions for entropy production and activity rates.
To prevent confusion, we note that the schematic of Fig.~\ref{fig:cartoon} was meant to depict generic joint distributions for activity and entropy production and does not illustrate $\psi_{\omega, K}(\lambda, s)$ for this particular solved model.

\section{Properties of the phase transition}
\label{sec:props}
\subsection{Tilted Operator Eigenvectors}
Thus far we have merely asserted that the trajectories have localized and delocalized character in the two dynamic phases, but here we more formally make the claims by analyzing the maximal right eigenvectors of the tilted operator.
We write the elements of the maximal eigenvector as $(f_1, f_2, \hdots, f_N)$ and note that the eigenvalue equation implies a recursion relation between neighboring $f_i$'s in the bulk.
\begin{equation}
\nonumber \begin{pmatrix}
f_i\\
f_{i+1}
\end{pmatrix} =
\begin{pmatrix}
\frac{\psi + 1 + x}{e^{-s} x^{1-\lambda}} & -x^{2\lambda - 1}\\
1 & 0
\end{pmatrix}
\begin{pmatrix}
f_{i+1}\\
f_{i+2}
\end{pmatrix}\\
=
B \begin{pmatrix}
f_{i+1}\\
f_{i+2}
\end{pmatrix},
\label{eq:Bdef}
\end{equation}
where we have introduced the transfer matrix $B$ and suppressed the subscripts and arguments on $\psi_{\omega, K}(\lambda, s)$.
The $n^{\rm th}$ component of the eigenvector can thus be written in terms of the two eigenvalues of of $B$, $k_1$ and $k_2$.
Specifically,
\begin{equation}
f_n \propto \left(k_1^{-1}\right)^n + \epsilon k_2^{(N-n)},
\label{eq:generaleigenvector}
\end{equation}
where $k_1 > 1$ and $k_2 < 1$.
The parameter $\epsilon$ serves to match up the boundary conditions between $f_1$ and $f_N$.
When $\lambda < \lambda^*$, the eigenvalues of $B$ can be expressed as $k_1 = e^{-\gamma / N}$, and $k_2 =x^{2\lambda - 1}e^{\gamma / N}$ correct up to second order in $1 / N$.
An expression for $\gamma$ in terms of the rate constants, Eq.~\eqref{eq:gamma}, follows from the full calculation of the maximum eigenvalue in Appendix~\ref{app:jointcalc}.
Hence the maximal right eigenvector is found to have components
\begin{equation}
f_n \propto e^{\gamma n / N} + \epsilon_{\rm deloc} e^{((2\lambda - 1) N \ln x + \gamma)(N-n)},
\label{eq:righteigenvectordelocalized}
\end{equation}
where $\epsilon_{\rm deloc} = h_1^{\lambda - 1}h_2^{-\lambda} (x^{1-\lambda} + h_2 - 1) - e^{\gamma}$.
The first term in Eq.~\eqref{eq:righteigenvectordelocalized} decays slowly over the entire range of the system, giving the eigenvector a delocalized character.
This delocalized character indicates that trajectories with high rates of entropy production and activity can be found regularly visiting all of the states of the system.

Under the conditions that $s > s^*(\lambda)$ the expression for $\gamma$ diverges, and the delocalized form for the maximal eigenvector given in Eq.~\eqref{eq:righteigenvectordelocalized} must break down~\footnote{Note that the cusp in the scaled cumulant generating function does not result from a simple eigenvalue crossing nor an avoided crossing as the delocalized vector ceases to be an eigenvector on the wrong side of the transition.}.
We anticipate a similar functional form for the eigenvectors except with some nonzero $\kappa$ replacing $\gamma / N$.
Indeed, with some tedious algebra it can be shown that the right eigenvectors are given by
\begin{equation}
f_n \propto e^{\kappa n} + \epsilon_{\rm loc} e^{((2 \lambda - 1) \ln x + \kappa)(N-n)}
\label{eq:righteigenvectorlocalized}
\end{equation}
with $\kappa = (\lambda^* - \lambda) \ln x$ and $\epsilon_{\rm loc} = h_1^{\lambda -1} h_2^{-\lambda}\left(x^{1-\lambda^*}+h_2-1\right)e^\kappa$.
Unlike the case of the delocalized eigenvector, this solution is strongly localized around the heterogeneous link.
Thus the flat region between the two cusps in $\psi_\omega(\lambda) \equiv \psi_{\omega, K}(\lambda, 0)$ stems from a class of localized trajectories which are incapable of producing entropy in the long-time limit.
The two dynamic phases can therefore be thought of as the classes of localized and delocalized trajectories, each of which contributes its own feature to the rate function.

\subsection{Entropy Production Statistics}
We previously reported on the entropy production statistics in the special case that $h_1 = h_2$~\cite{Vaikuntanathan2014}, but this restriction was lifted in the preceding analysis.
Surprisingly, allowing for distinct values of $h_1$ and $h_2$ can yield a qualitative difference in the entropy production rate statistics.
When $h_1$ and $h_2$ are constrained to be equal, all values of $h$ and $x$ give rise to singularities in $\psi_\omega(\lambda)$ and therefore a dynamic phase transition with respect to the entropy production rate.
By solving for the conditions when $s^*(\lambda) = 0$ in Eq.~\eqref{eq:sstar} one can obtain the position of these two cusps in $\psi_{\omega}(\lambda)$, $\lambda^*$ and $1-\lambda^*$.

\begin{figure}[tbhp]         
\centering
\includegraphics[width=0.98\linewidth]{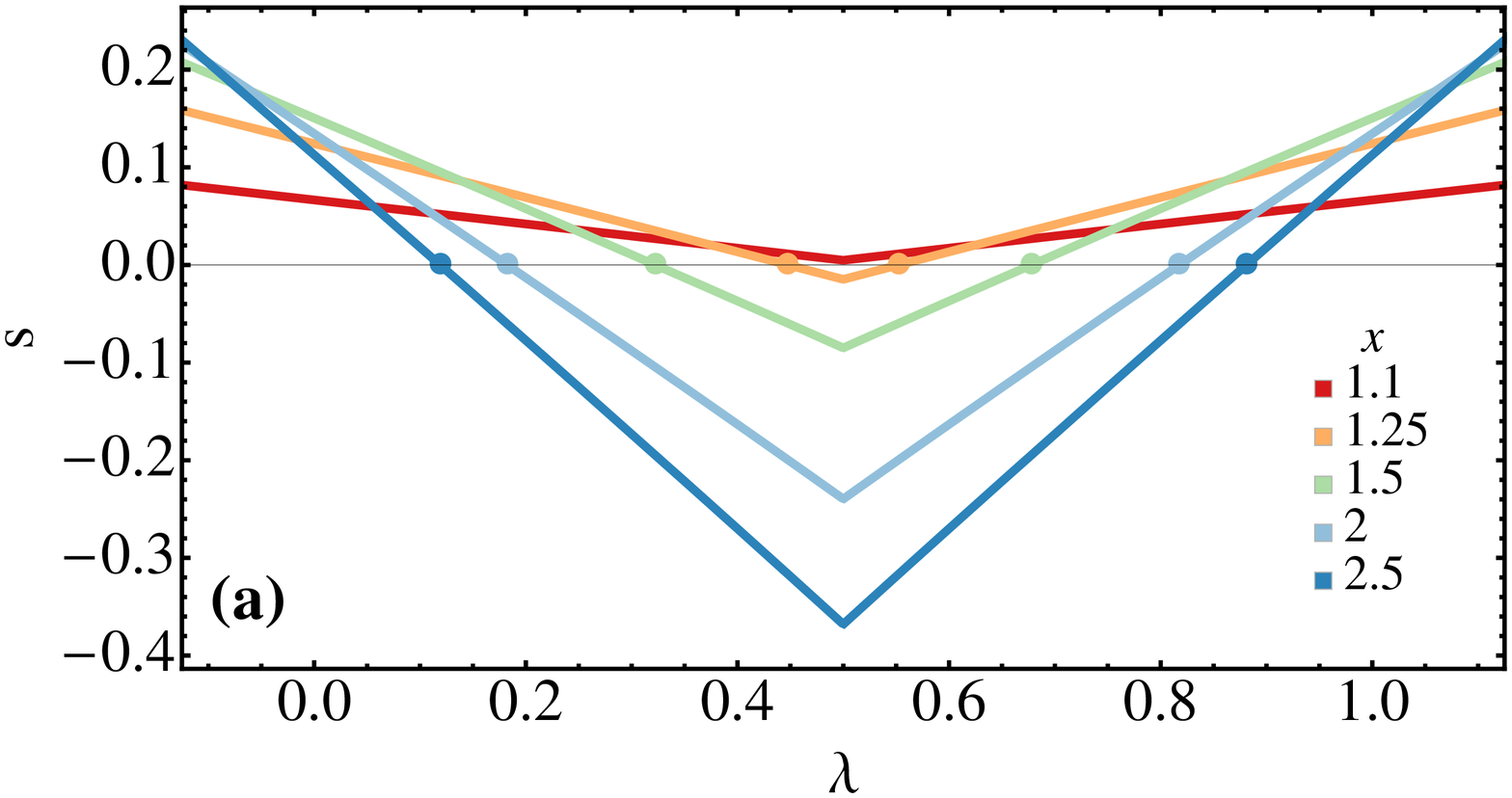}
\includegraphics[width=0.98\linewidth]{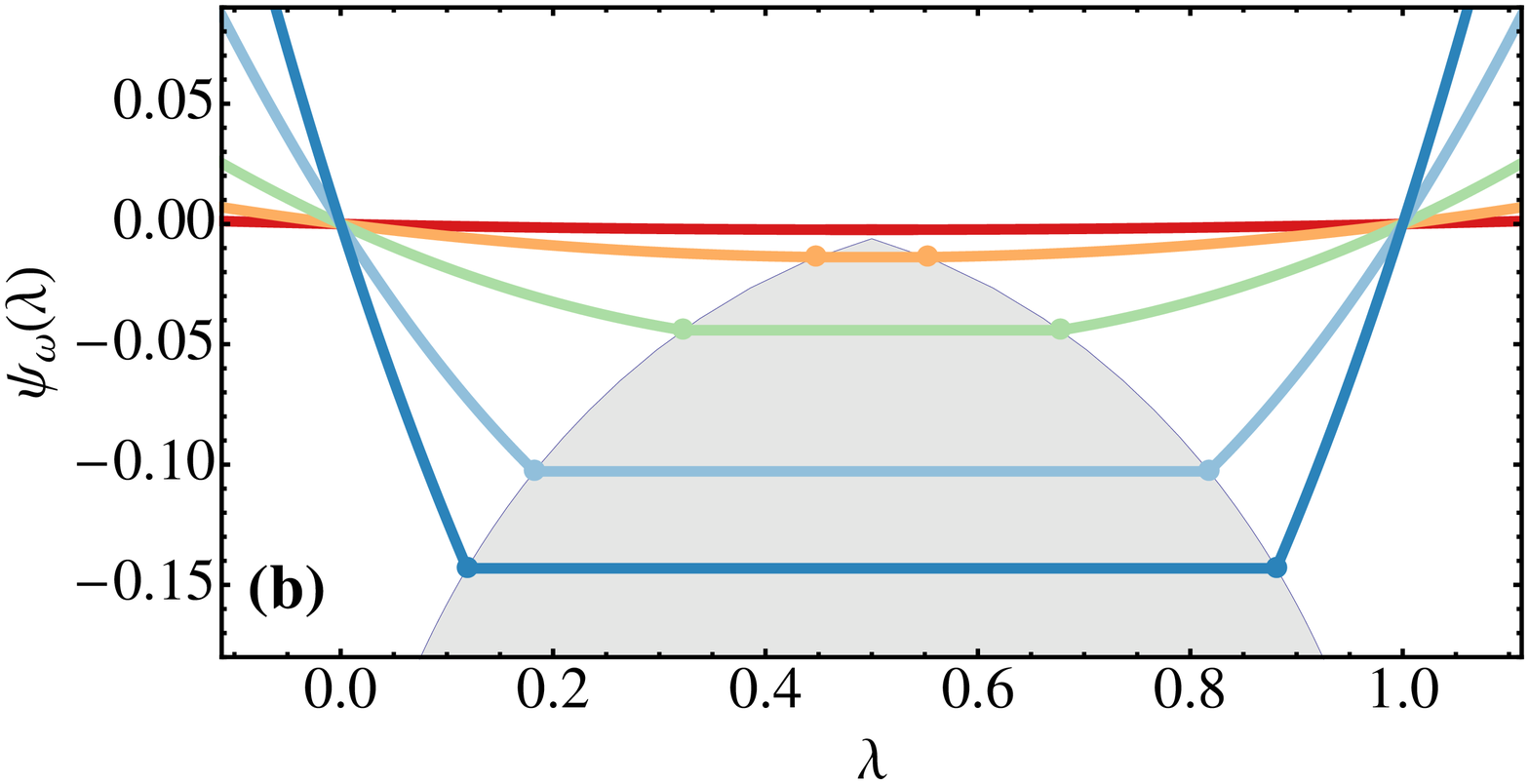}
\caption{(Color online) (a) Curve of the cusp in $\psi_{\omega, K}(\lambda, s)$ given by Eq.~\eqref{eq:sstar} plotted for $h_1 = 0.3$ and $h_2=0.2$.
This curve corresponds to the dashed line in the schematic of Fig.~\ref{fig:cartoon}(f). 
Below the curve $\psi_{\omega, K}(\lambda, s)$ has a delocalized eigenvector. 
For these parameters $x_c \approx 1.163$ such that the $x = 1.1$ curve does not intersect the $s=0$ axis.
(b) $\psi_\omega(\lambda)$ for the same conditions.
Cusps are marked with a filled dot at $(\lambda^*, \psi_\omega(\lambda^*))$ and $(1-\lambda^*, \psi_\omega(1 - \lambda^*))$.
The shaded area indicates the region arising from the localized phase.
}
\label{fig:psih1h2xvals}
\end{figure}

When $h_1$ and $h_2$ are distinct, however, there are conditions for which $s^*(\lambda)$ lacks roots. 
The fluctuation theorem~\cite{Lebowitz1999} entails a symmetry in the cumulant generating function $\psi_{\omega}(\lambda)$ about $\lambda=1/2$. 
Because of this symmetry, the marginal case where the dynamic phase transition disappears occurs when $\lambda^* = 1/2$.
Solving for the condition that $s^*(1/2) = 0$ thus gives a critical value of $x$,
\begin{align}
\nonumber x_\text{c} = \frac{1}{2} \bigg(1 + 2h_1 &- 2h_2 + h_2^2 -\\
& (h_2 - 1)\sqrt{1 + 4h_1 - 2h_2 + h_2^2} \bigg),
\label{eq:xcritical}
\end{align}
so that $\psi_\omega(\lambda)$ will have cusps indicating a dynamic phase transition if and only if $x > x_{\rm c}$.
Indeed, Fig.~\ref{fig:psih1h2xvals} illustrates that the singularities are no longer present when $x$ drops below the critical value $x_{\rm c}$.
It is of particular note that the critically does not occur at the trivial limit $x = h_1 / h_2$, the condition for which hops across the heterogeneous link produce the same amount of entropy as hops across any other link.

\begin{figure*}[tbhp]         
\centering
\includegraphics[width=0.48\linewidth]{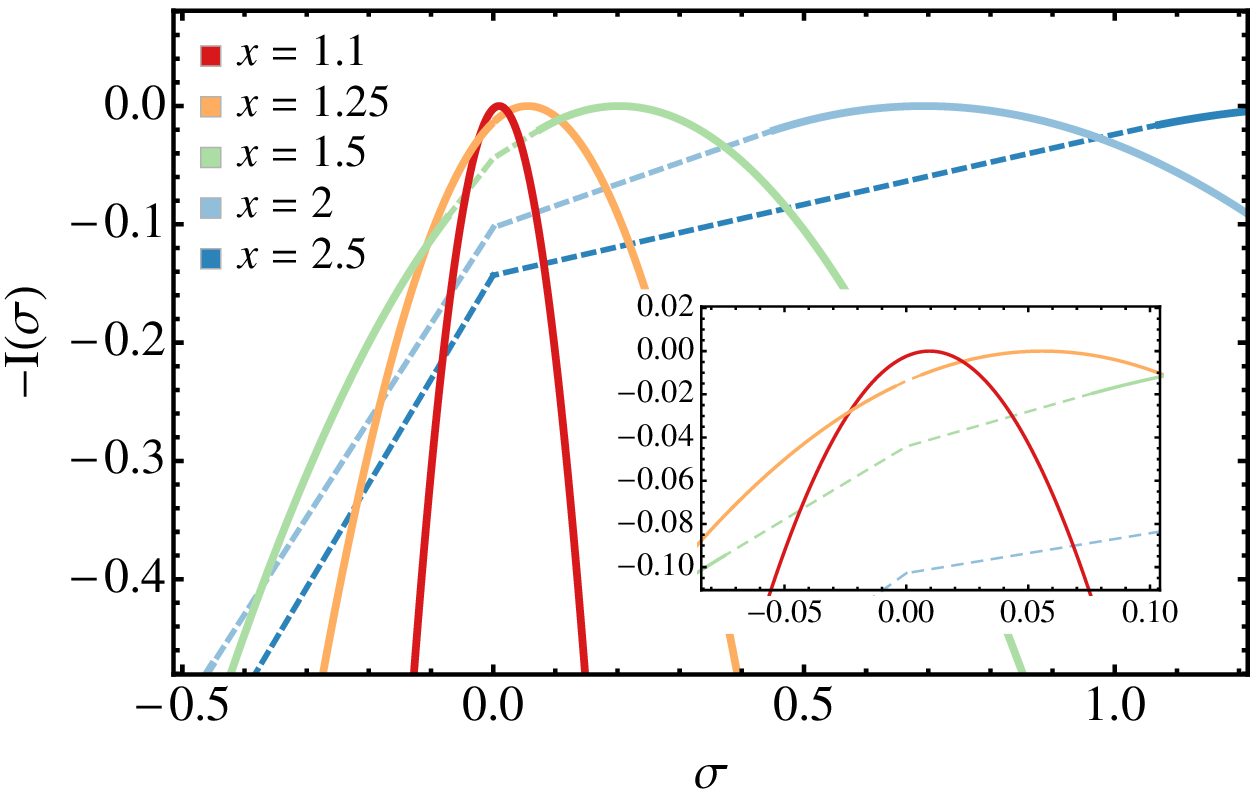}
\hspace{0.02\linewidth}
\includegraphics[width=0.48\linewidth]{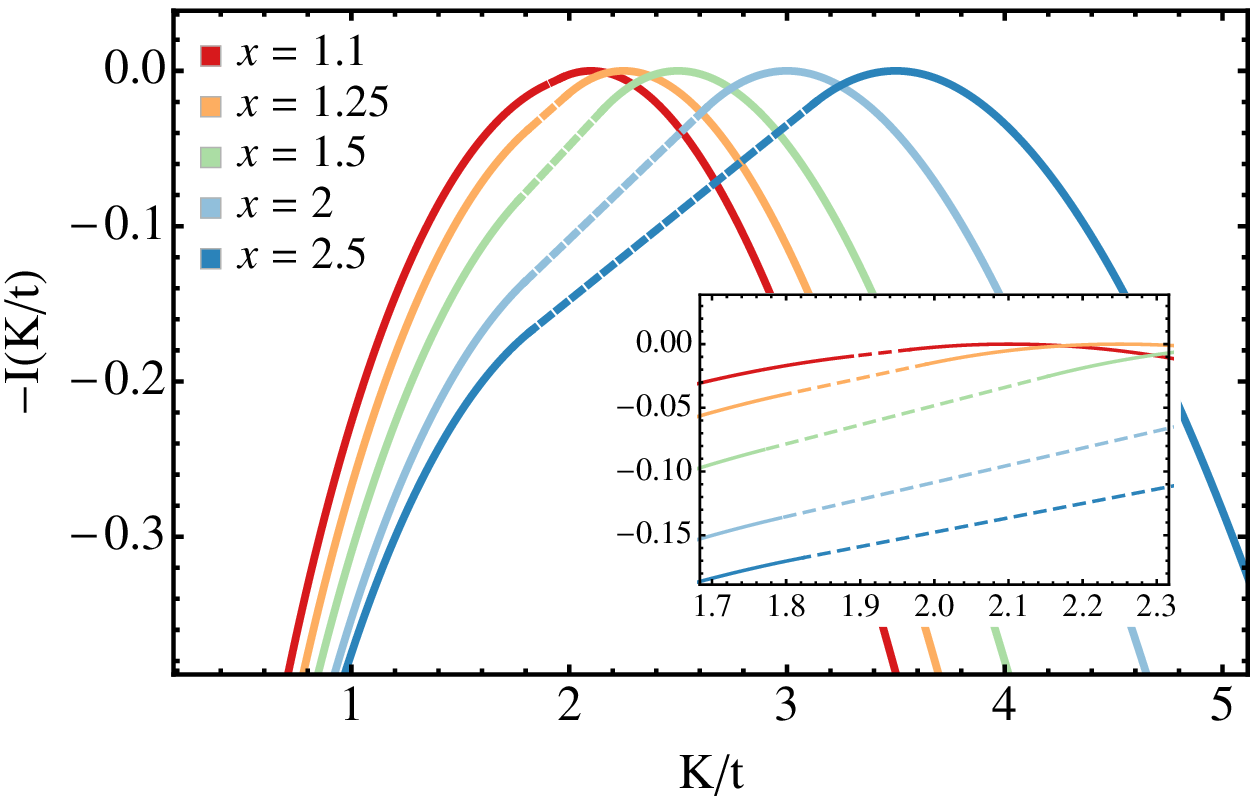}
\caption{(Color online) Entropy production and activity rate functions for $h_1 = 0.3$ and $h_2 = 0.2$ where $x_{\rm c} \approx 1.163$.
The insets zoom in on the peaks to illustrate that the entropy production rate function lacks a second phase (and consequently does not have the dashed tie lines), while the activity rate function shows two phases even for $x < x_{\rm c}$. The most likely values of $\sigma$ and $K/t$ increase with increasing $x$.
}
\label{fig:xvalsplots}
\end{figure*}

The existence of a critical value of $x$ can be understood more clearly by examining the large deviation rate function for the entropy production, $I(\sigma)$, which is obtained from a Legendre transform of $\psi_{\omega}(\lambda)$ .
In Fig.~\ref{fig:xvalsplots} we plot these rate functions for a variety of values of $x$ but for the same value of $h_1$ and $h_2$.
As $x$ is decreased, the system is biased less strongly toward clockwise cycles, and the average entropy production rate decreases correspondingly.
However, even when the average entropy production rate is large, the class of localized trajectories present a way for a trajectory to produce zero entropy.
Therefore a broad entropy production distribution with a hump at $\sigma = 0$ is present for large $x$.
When $x$ is decreased below $x_c$, this shoulder at $\sigma = 0$ gets completely engulfed by the natural fluctuations in entropy production characterizing the dominant (delocalized) class of trajectories.
Thus the disappearance of the dynamic phase transition corresponds to the condition when near-zero entropy production rates are more likely to be obtained by a delocalized trajectory than by a localized trajectory.
As we shall demonstrate shortly, the lack of the dynamic phase transition in the entropy production order parameter does not rule out the presence of two classes of trajectories. A dynamic phase transition can still be recovered by studying the statistics of dynamical activity. 

\subsection{Dynamical Activity Statistics}
The statistics for the dynamical activity can be deduced in the same way by setting $\lambda = 0$ in Eq.~\eqref{eq:jointcgf}.
The activity does not satisfy a fluctuation theorem, so there is no symmetry corresponding to $\psi_\omega(\lambda) = \psi_\omega(1-\lambda)$.
As a consequence, $\psi_{K}(s) \equiv \psi_{\omega, K}(0, s)$ has a cusp at $s^*(0)$ for all values of the rate constants as is clear from the $\lambda = 0$ intercepts of Fig.~\ref{fig:psih1h2xvals}.

The Legendre transform of $\psi_K(s)$ gives the rate function for dynamical activity, shown in Fig.~\ref{fig:xvalsplots}.
Like the case of entropy production, as $x$ is decreased the average activity decreases, but now the tie line (and correspondingly the dynamic phase transition) persists for all choices of $x$.
Remarkably, this implies that there is a regime with $x < x_{\rm c}$ where the activity exhibits a dynamic phase transition but the entropy production does not.
Consequently the entropy production distribution in this parameter regime converges to the distribution found in a translationally symmetric network, while the distribution for dynamical activity resolves the impact of the heterogeneity.

\begin{figure*}[tbhp]         
\centering
\includegraphics[width=0.28\linewidth]{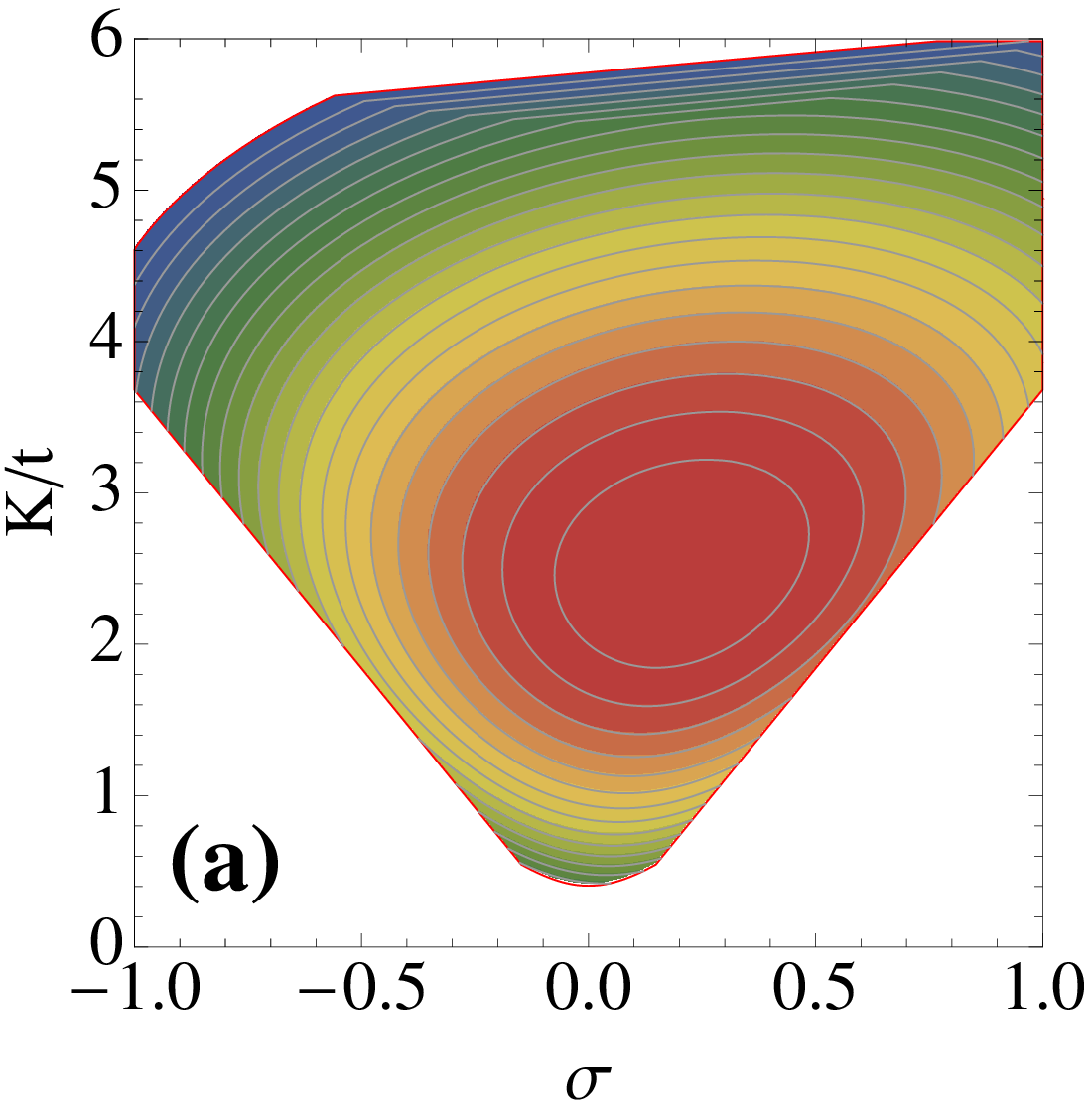}
\includegraphics[width=0.28\linewidth]{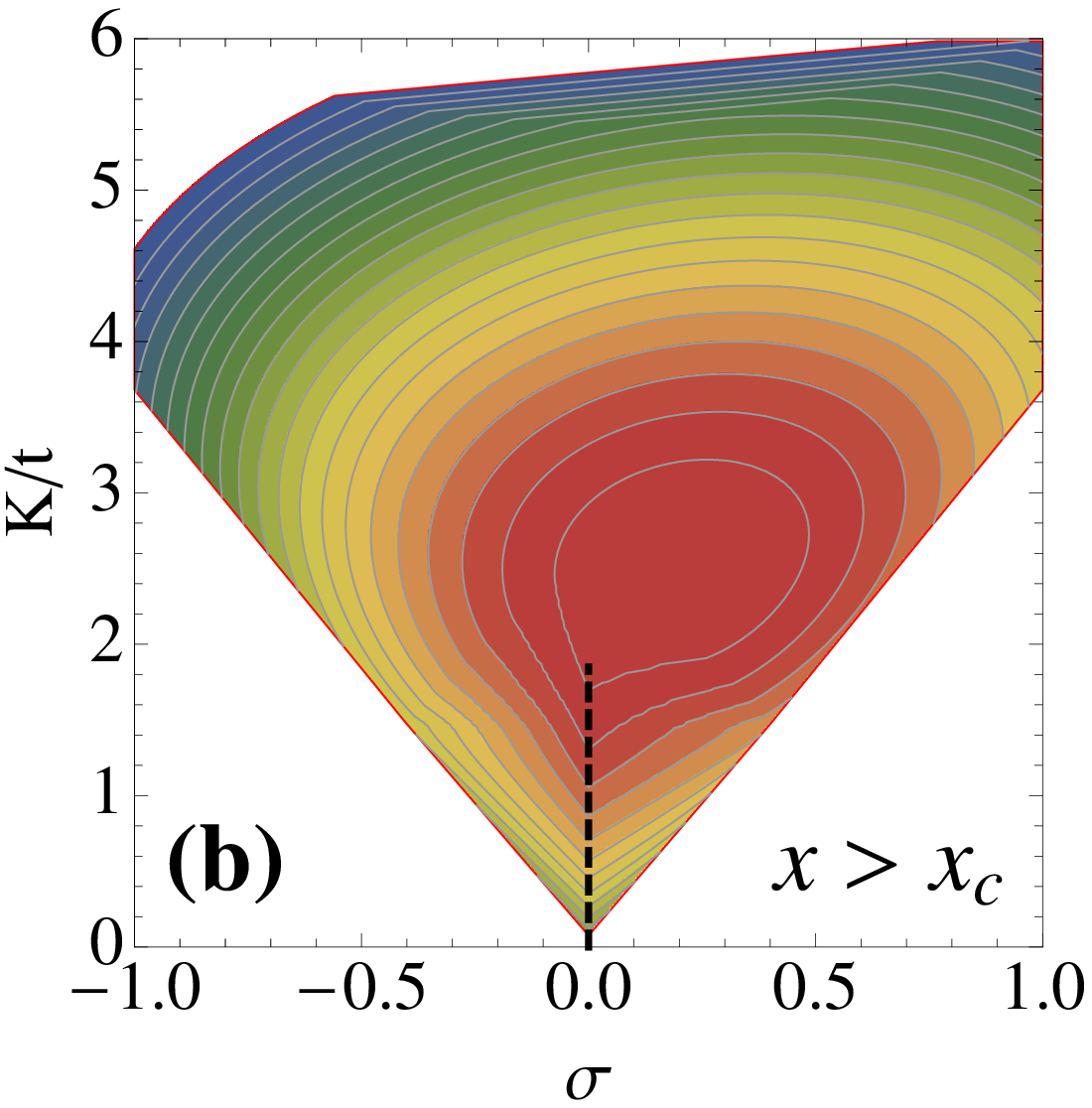}
\includegraphics[width=0.28\linewidth]{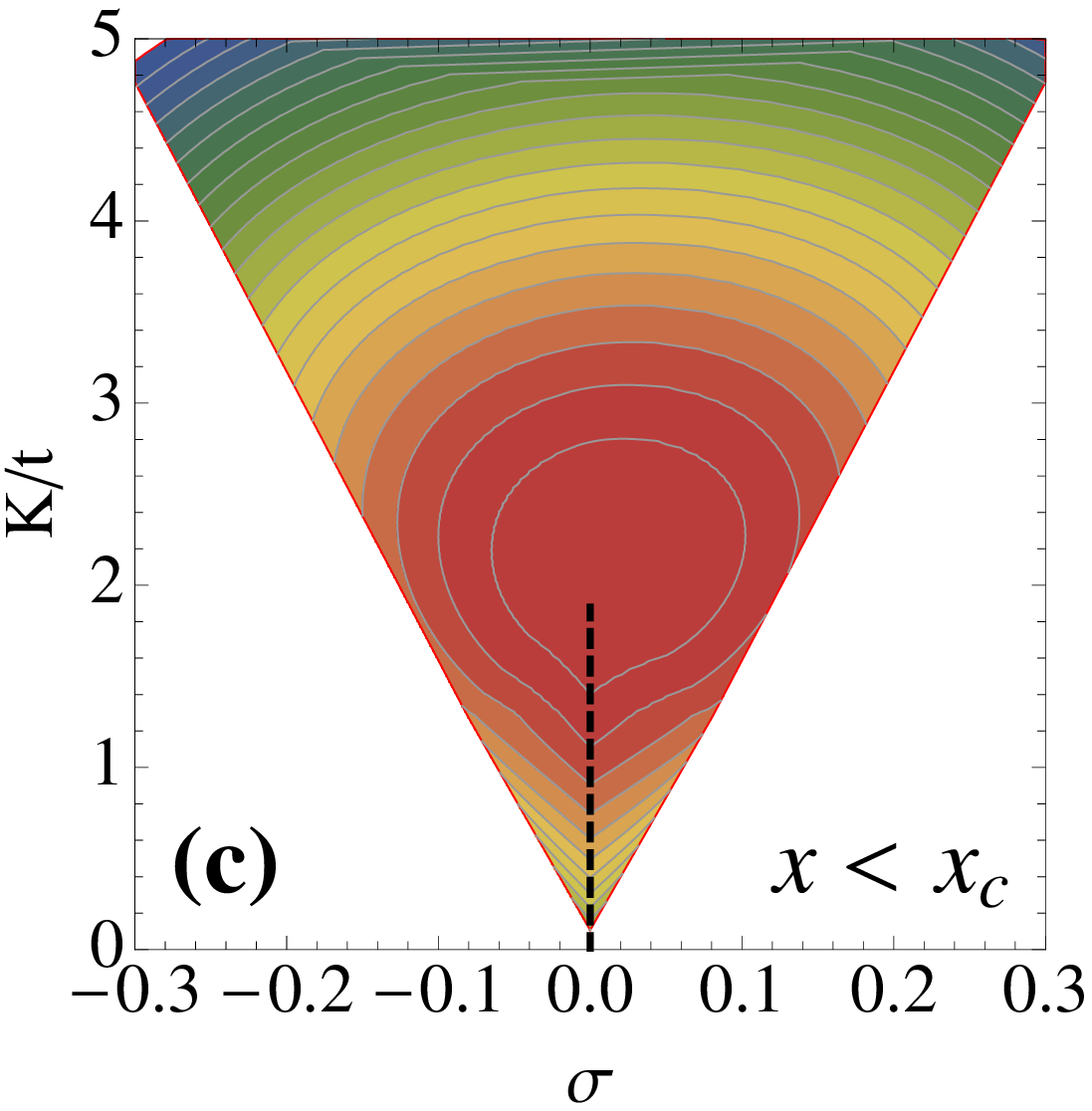}
\includegraphics[width=0.08\linewidth]{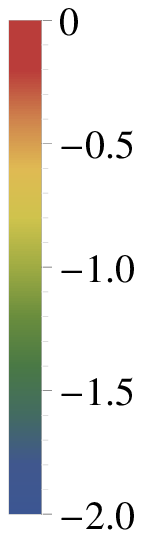}\\
\includegraphics[width=0.28\linewidth]{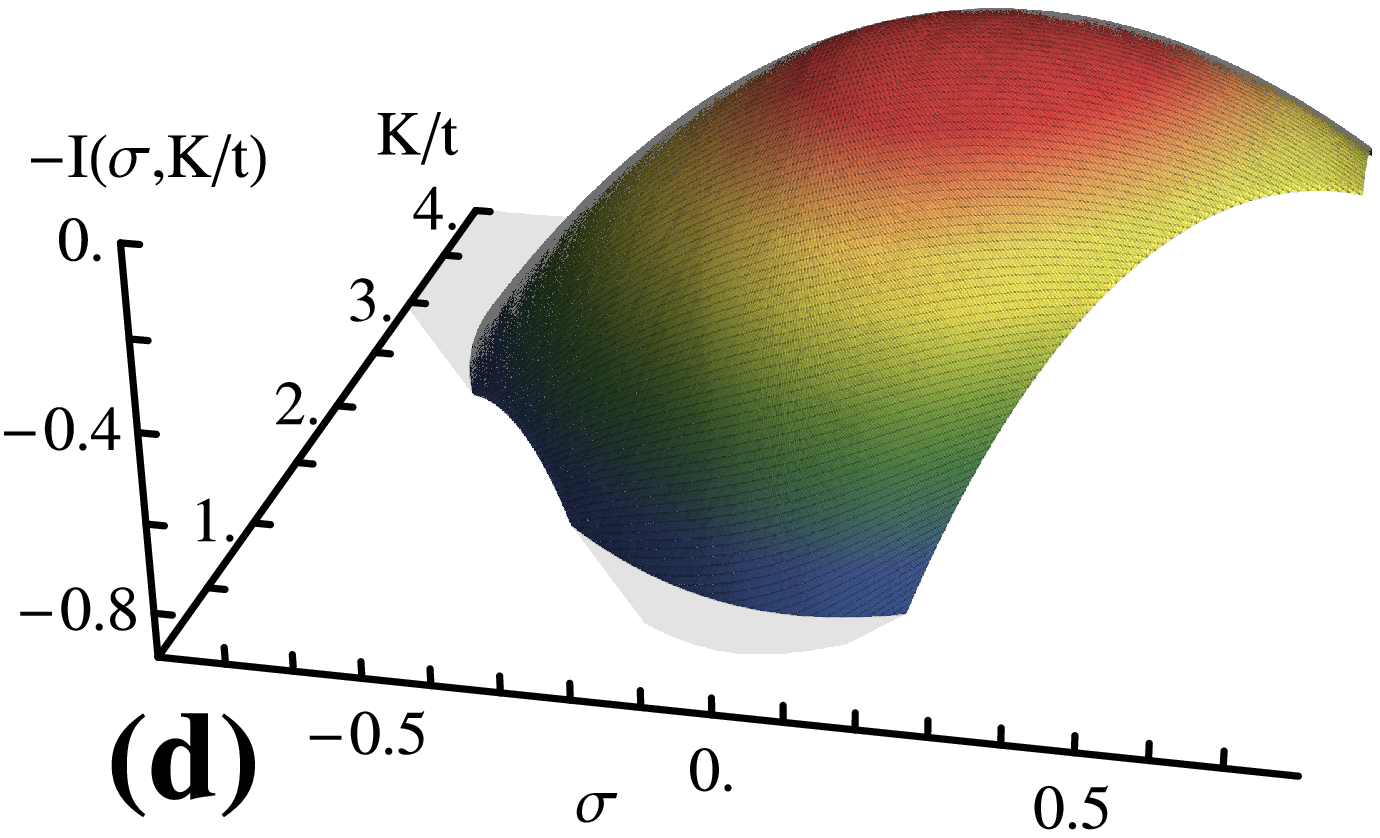}
\includegraphics[width=0.28\linewidth]{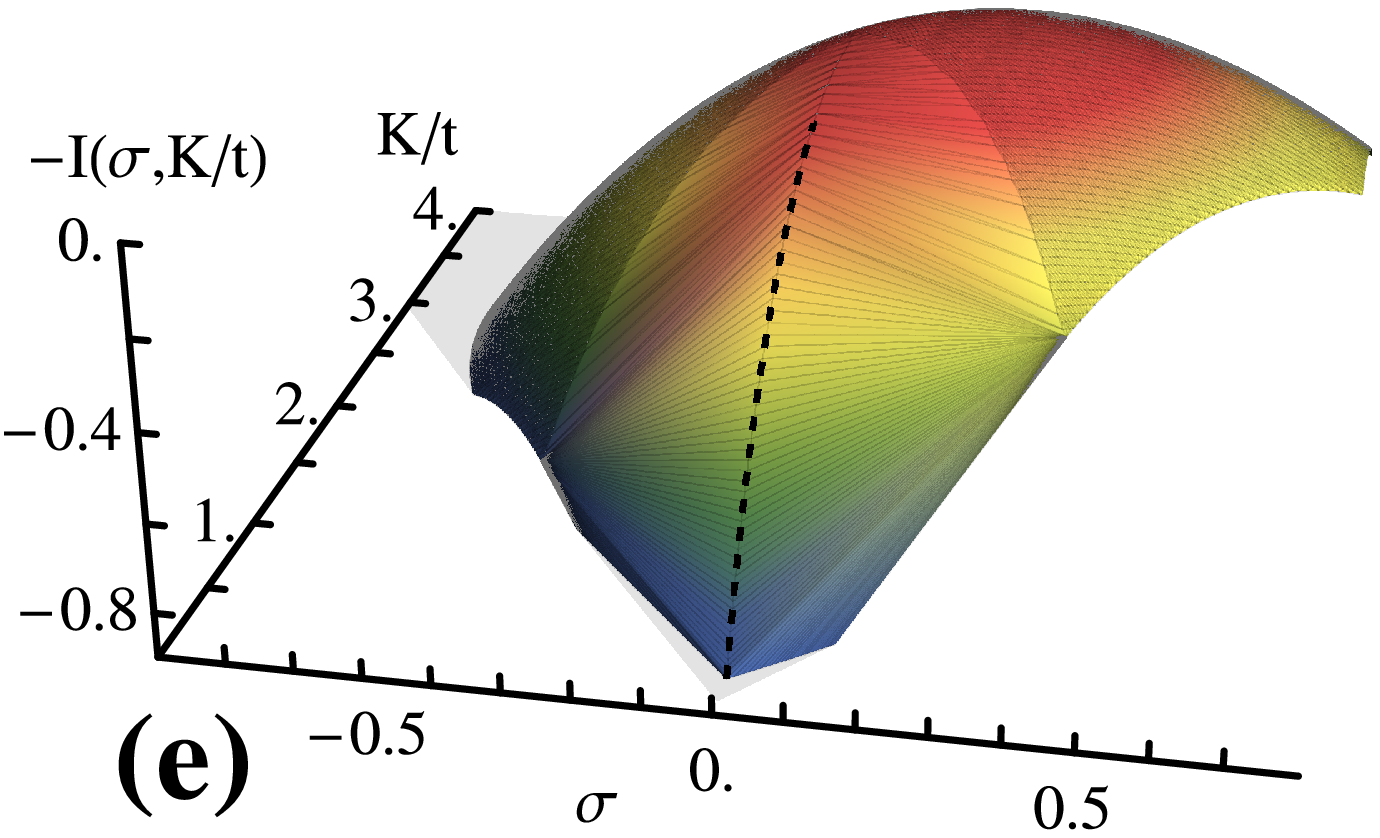}
\includegraphics[width=0.28\linewidth]{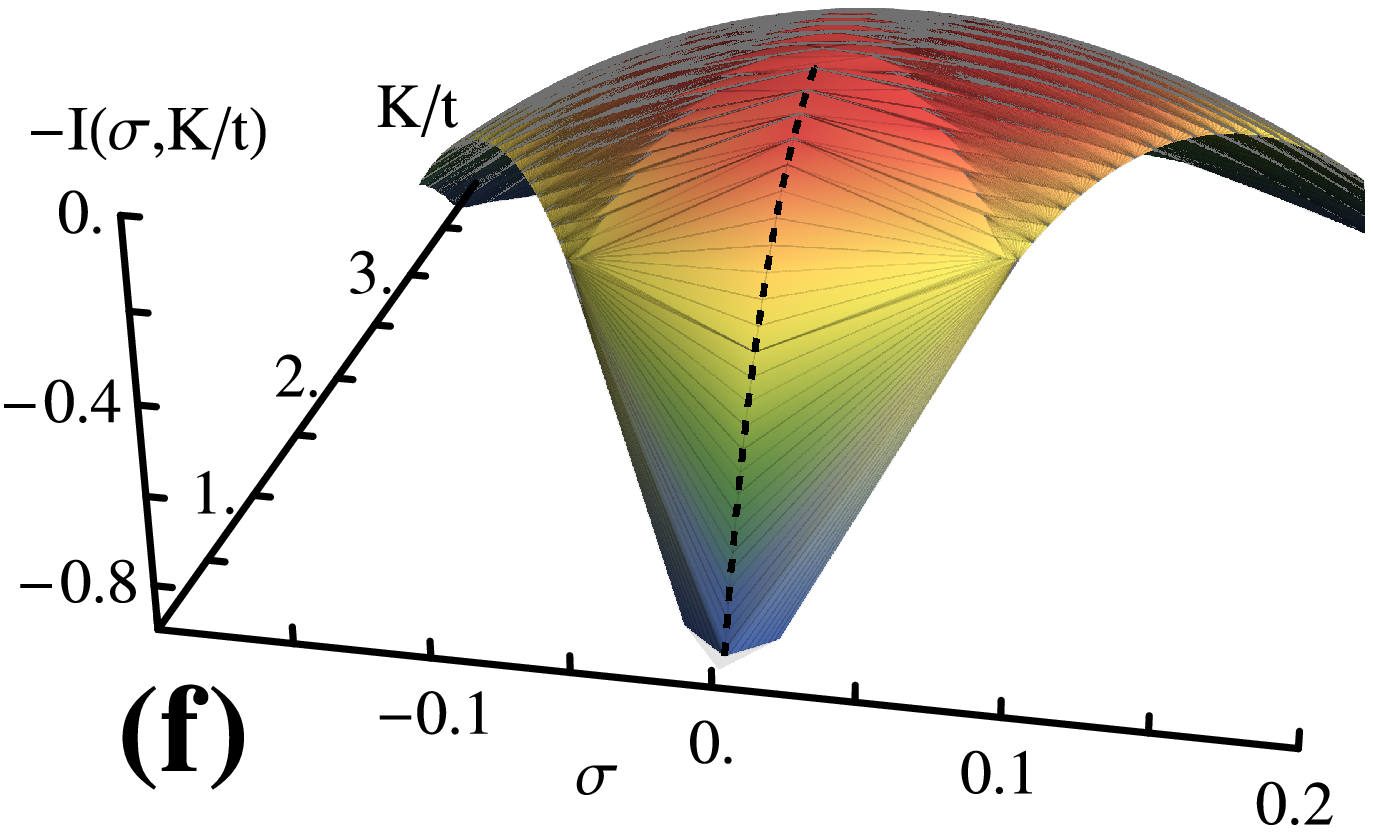}
\hspace{0.08\linewidth}
\caption{(Color online) Joint rate functions for the activity and entropy production rates. 
The translationally symmetric case with $x=1.5$ and no heterogeneity is plotted in (a) and (d).  
The influence of a heterogeneous link ($h_1 = 0.3, h_2 = 0.2$) is shown in (b) and (e) for $x = 1.5$ and in (c) and (f) for $x = 1.14$.
In the presence of the heterogeneity, the rate function develops a set of tie lines to connect two dynamic phases.
The localized phase runs along the $\sigma = 0$ ridge (dashed line), enabling fluctuations in $K/t$ but not in $\sigma$.  
The delocalized phase is centered at the peak of the rate function and enables fluctuations in both $K/t$ and $\sigma$.
Whether $x$ exceeds $x_{\rm c}$ determines if localized phase is visible to the entropy production axis.
}
\label{fig:jointratefunctions}
\end{figure*}
To better appreciate the manner in which a dynamic phase transition can be observed with respect to one order parameter but not another, we Legendre transform $\psi_{\omega, K}(\lambda, s)$ and plot the two dimensional rate functions for activity and entropy production.
Figures~\ref{fig:jointratefunctions}(a) and (d) show the rate function for the translationally symmetric network with $x = 1.5$.
Note that the surface is smooth, exhibiting only small fluctuations away from the mean behavior.
In contrast, when the heterogeneous link is introduced a ridge develops along $\sigma = 0$.  
This ridge corresponds to the class of localized trajectories, all of which have identically zero entropy production rate in the long time limit.
The entropy production and activity rate functions are marginals of this two dimensional surface, which corresponds to projecting the surface onto the $\sigma$ and $K / t$ axes, respectively.
For all values of $x$, the projection onto the $K / t$ axis results in a broad activity distribution with components from both the localized and delocalized trajectories.
The projection onto the $\sigma$ axis behaves differently.  
When $x < x_{\rm c}$ the class of localized trajectories along the ridge are in line with the most likely contributions from delocalized trajectories.
Consequently the entropy production distribution will not reveal the localized trajectories since for all possible values of $\sigma$ there exist more probable delocalized trajectories which produce that particular entropy production rate.

The calculation offers an important lesson which provides insight for more complicated dynamical systems.
Our analysis has shown that localized and delocalized trajectories can be clearly separated into two distinct classes.
Nevertheless, the underlying transition is only visible in the distribution for certain order parameters.
In more complicated systems, one can expect many more than two classes of trajectories.
Whether or not these classes constitute a true dynamical phase is intimately related to the symmetries of the dynamic order parameter being probed.
Thus an experimenter simultaneously monitoring current, activity, and entropy production distributions may consistently observe large deviations in some order parameters but not in others. 
We note that similar scenarios can occur in equilibrium statistical mechanics. 

\section{Physical Implications of Dynamical Phases}
\subsection{Tuning Rates to Coexistence}

We have shown that while typical trajectories on the network are delocalized, there exists a rare class of localized trajectories, which in certain cases appears as a distinct dynamic phase.
In ordinary statistical mechanics one tunes a Lagrange multiplier like the inverse temperature, $\beta$, to induce a transition between phases.
Inducing a dynamic transition cannot occur in an identical way since the $\lambda$ field is conjugate to a time-non-local object.
In computer simulations one can place the system in contact with a large bath at a well-defined value of $\lambda$, but this treatment requires that the system is the entire trajectory.
While it is not possible to directly tune the $\lambda$ field in an experiment, one can use the eigenvectors of the tilted operator to construct a rate matrix which, in the long-time limit, is equivalent to introducing such a $\lambda$ field~\cite{Jack2010, Chetrite2013}.
We denote this class of rate matrices for effective $\lambda$ fields $\mathcal{W}(\lambda)$ with matrix elements given by
\begin{equation}
\mathcal{W}_{ij}(\lambda) = \frac{f_i(\lambda)}{f_j(\lambda)} \left[\mathbb{W}_\omega(\lambda)\right]_{ij} - \delta_{ij} \psi_\omega(\lambda),
\label{eq:effectivematrix}
\end{equation}
where $f_n$ is the $n^{\rm th}$ component of the right eigenvector determined in the large $N$ limit in Eqs. \eqref{eq:righteigenvectordelocalized} and \eqref{eq:righteigenvectorlocalized}.
For simplicity we limit the analysis to the $\lambda$ field conjugate to entropy production, but this could of course be repeated for activity. 
\begin{figure}
\centering
\includegraphics[width=0.88\linewidth]{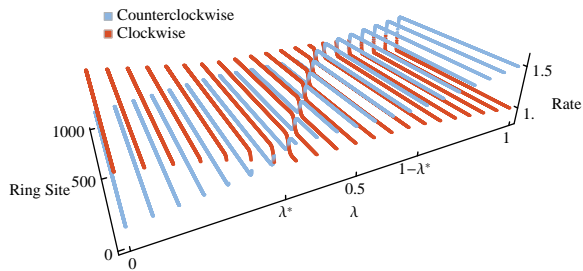}
\caption{(Color online) The off-diagonal elements of $\mathcal{W}(\lambda)$ give effective rate constants for clockwise and counterclockwise transitions rates after the application of a $\lambda$ bias.  
In the absence of $\lambda$ biasing the transition rates are $h_1 = 0.3$, $h_2 = 0.2$, $x = 1.5$. 
Note the strong spatial dependence of the effective rate constants between $\lambda^*$ and $1- \lambda^*$ where the typical trajectories are localized.}
\label{fig:lambdarates}
\end{figure}

A cusp in the maximum eigenvalue of the tilted operator $\mathbb{W}_\omega(\lambda)$ indicates that the two dynamic phases will be in coexistence when a $\lambda^*$ bias is applied.
The tilted operator, however, is only a rate matrix at $\lambda = 0$.
For all other values of $\lambda$ the matrix does not conserve probability. 
In contrast, $\mathcal{W}$, is a proper rate matrix and can consequently represent physical rates which result in long-time dynamics that exactly mimics the long-time behavior of the original dynamics subject to a $\lambda^*$ biasing field.~\cite{Jack2010,Chetrite2013}.
For our solvable model, we determine the clockwise and counterclockwise rate constants as a function of position around the ring which yield an effective $\lambda$ bias according to Eq.~\ref{eq:effectivematrix}.
These rates are depicted in Fig.~\ref{fig:lambdarates}.
When $\lambda = 0$ they are just the rate constants of the natural dynamics, but as $\lambda$ increases the preference toward low entropy production has two effects.
Firstly, the effective clockwise rates decrease while counterclockwise rates increase to yield slower rates of cycling around the ring.
With fewer completed cycles per unit time, the trajectories achieve a lower rate of entropy production.
Secondly, a strong spatial dependence in the rates arises, particularly as $\lambda$ nears $\lambda^*$.
As long as the clockwise rates exceed the counterclockwise rates at every site in the ring, typical trajectories will remain delocalized.
In the regime $\lambda^* < \lambda < 1-\lambda^*$ some regions of the ring prefer clockwise motion while others prefer counterclockwise.
The result is a localization at the interface of these regions.
As made clear in Fig.~\ref{fig:lambdarates}, the effective rates are smoothly tuned by $\lambda$, even when passing through the transition.
Dynamics evolving under $\mathcal{W}(\lambda^*)$ is particularly interesting as it will exhibit massive fluctuations in entropy production rates.
Given that states are connected together in a ring topology, this $\mathcal{W}(\lambda^*)$ gives the physical rate constants which tune the system to dynamical coexistence.

The ring topology is also convenient for demonstrating that $\lambda$ biasing of detailed balance systems results in dynamics which also obeys detailed balance.
This has previously been shown to be the case when biasing activity by an $s$ field by other means~\cite{Jack2010}.
We note that in a network with a cycle loop (like our ring network) the condition of detailed balance is satisfied if and only if 
\begin{equation}
\prod_{i=1}^{N} \left(\frac{\mathbb{W}_{i,i+1}}{\mathbb{W}_{i+1,i}}\right) =1
\label{eq:db}
\end{equation}
i.e.\ if the product of rates for clockwise transitions equals the product of rates for counterclockwise transitions~\cite{Schnakenberg1976}.
(We have implicitly assumed periodic boundary conventions.)
Constructing similar products for effective dynamics under a $\lambda$ field (Eq.~\ref{eq:effectivematrix}), we find that the effective dynamics obey detailed balanced if and only if
\begin{equation}
\prod_{i=1}^{N} \left(\frac{\mathbb{W}_\omega(\lambda)_{i,i+1}}{\mathbb{W}_\omega(\lambda)_{i+1, i}}\right) = 1.
\label{eq:db2}
\end{equation}
Using the definition of $\mathbb{W}_\omega(\lambda)$, in Eq.~\ref{eq:Woperatordef}, Eq.~\ref{eq:db2} can be expressed as
\begin{equation}
\left(\prod_{i=1}^{N} \frac{\mathbb{W}_{i,i+1}}{\mathbb{W}_{i+1,i}}\right)^{1-2\lambda} =1\,.
\label{eq:db4}
\end{equation}
In other words, provided $\lambda \neq 1/2$, the effective dynamics satisfy the condition of detailed balance if and only if the underlying physical dynamics are detail balanced. 
The result is simply extended to networks with multiple cycles using the cycle decomposition theorem~\cite{Schnakenberg1976}.

\subsection{Persister Cells}
It has long been observed that a small fraction of a colony of genetically identical bacterial cells are resistant to antibiotic treatment~\cite{Bigger1944}. 
One important observation is that bacteria which are not dividing are not affected by the antibiotic, which suggests that bacteria have an internal switch allowing rare transitions into non-dividing persister states that could provide protection from the antibiotic~\cite{Lewis2010}. 
Several detailed mechanisms have been proposed for stabilizing the non-dividing persister state of the bacteria ~\cite{Lewis2000, Keren2004}, though these pathways have been shown to be not wholly responsible for the appearance of persister cells ~\cite{Vazquez2006}.

We note that the dynamic phase transition of our studied model presents a distinct stochastic hypothesis to explain the long timescale decay of bacterial population in response to an antibiotic.
The ensemble of bacteria could be thought of as the ensemble of trajectories evolving in time around the ring, with every completed cycle corresponding to another cell division.
While typical cells cycle rapidly, a rare dynamical phase of localized, non-dividing cells could be expected to exist solely because of the heterogeneity of rates around a cell cycle.
Provided that antibiotics kill cells which grow rapidly, the localized subensemble of cells predicted by our calculations could result in an anomalously slow decay in survival probability.
Notably, our model lacks an explicit degree of freedom capable of differentiating persister and normal states based on a single-time observation since the phases describe classes of trajectories, not of configurations. 

\begin{figure}[tbhp]         
\centering
\includegraphics[width=0.88\linewidth]{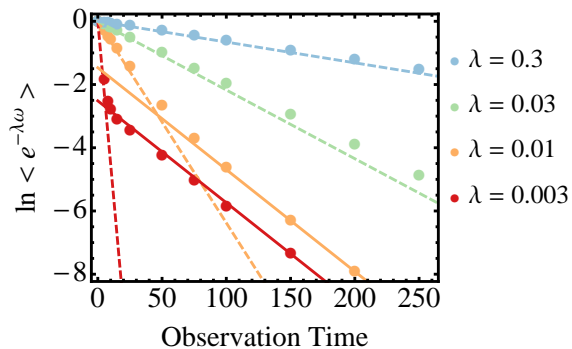}
\caption{(Color online) Log survival probability for trajectories killed in proportion to $e^{-\lambda \omega}$.
Entropy production for $5 \times 10^6$ steady state trajectories of various observation times were collected with $x=3, h_1=h_2=0.05, N=1000$.
For this choice of rate constants, $\lambda^* \approx 0.015$.
At the observation time trajectories are killed with a probability tuned by $\lambda$, which could act like the strength of an antibiotic in the case of bacterial cells.
Dashed lines are lines with slope $\psi^{\rm(ts)}(\lambda)$, which capture the short-time behavior since typical trajectories do not encounter the heterogeneity at short times.
Solid lines have slope $\psi(\lambda)$ as given by Eqs.~ \eqref{eq:1Nexpansion} and \eqref{eq:gamma}.
As the maximum eigenvalue, $\psi(\lambda)$ must characterize the long-time behavior.
}
\label{fig:timescaling}
\end{figure}

In Figure~\ref{fig:timescaling} we use our model system to extract qualitative estimates of the survival probability of cells as a function of time in the presence of an antibiotic. 
In particular, we kill trajectories in proportion to $e^{-\lambda \omega}$, where $\omega$ denotes the entropy produced along the trajectory and $\lambda$ controls the death rate and is meant to represent the presence of an antibiotic. At long times, the log probability of surviving trajectories decays according to $\psi_\omega(\lambda)$, which differs from the initial decay rate if $\lambda > \lambda^*$ and the dynamic phase transition can be accessed. The decay of survival probability can be expected to change markedly from single exponential to biexponential behavior at a critical value $\lambda^*$.
An experimental realization of such an observation may be accessible in observing the survival probability of persister cells in response to different classes of antibiotics.

We note that the bi-exponential curves in Fig.~\ref{fig:timescaling} are similar to those observed experimentally~\cite{Balaban2004} and are attained even without an internal switch that determines the cell's type.
This demonstration suggests that persister cells may be a generic feature resulting from inescapable heterogeneity in transition rates, such that removing one pathway implicated in supporting persisters will just reveal new localized phases centered around different heterogeneous links.

\section{Conclusions}
We have investigated the properties of a dynamical phase transition in a recently introduced exactly solvable model. 
Using methods of large deviation theory we analytically computed the joint rate function for the dynamical activity and entropy production rates for a single-particle system evolving on a simple driven kinetic network. 
The joint rate function demonstrates two dynamical phases \textemdash \ one localized and the other delocalized \textemdash \ but the marginal rate functions do not exhibit the underlying transition under all conditions. 
Specifically, the marginal rate function corresponding to the entropy production has a critical point beyond which there is no dynamic phase coexistence even though the system still supports two distinct classes of trajectories. 
We illustrated the rates that position the system in a state of coexistence between localized and delocalized phases. 
We also discussed a biophysical implication of the transition, namely the heterogeneity in the growth rates of bacterial cells and the phenomena of persistence, arguing that a bistability of dynamical phases can be found in even the simplest models of a cell cycle which lack an explicit bistability in configuration space.

Our study of a single-particle system reveals one simple manner in which dynamical phase transitions can arise.
The two dynamical phases are time independent, a characteristic of the transition shared by some many-particle systems~\cite{Bodineau2004}, but the model certainly does not encompass the full range of dynamical complexity intrinsic to interacting nonlinear degrees of freedom.
This work thus does not directly help to clarify, for instance, rare phases related to the hydrodynamic limit of many-particle dynamics~\cite{Bodineau2005, Bertini2005, Bertini2006}.
Nevertheless, our exact analytical results for a schematic model highlight features that could be important in much more exotic phenomena, most notably the possibility that dynamic phase transitions can be visible to some pertinent order parameters yet hidden from others.

\section*{Acknowledgments}
T.R.G. acknowledges support from the NSF Graduate Research Fellowship and the 
Fannie and John Hertz Foundation. 
This work was supported in part by the Director, Office of Science, Office of Basic Energy Sciences, Materials Sciences, and Engineering Division, of the 
U.S. Department of Energy under contract No.\ DE AC02-05CH11231 
(S.V. and P.L.G.).

\appendix
\section{Computation of Scaled Cumulant Generating Function}
\label{app:jointcalc}
Were it not for the heterogeneous link connecting sites $1$ and $N$, the tilted operator would have a translational symmetry, making its explicit diagionalization trivial in a Fourier basis.
We construct a $1 / N$ expansion of $\psi_{\omega, K}(\lambda, s)$ by expanding around around the maximum eigenvalue of the translationally symmetric network,
\begin{align}
\nonumber \psi_{\omega, K}(\lambda, s) &= x^{1-\lambda}e^{-s} + x^\lambda e^{-s} - 1 - x \\
& \ \ \ \ \ \ \ \ \ \  \  \ \ \ \ \ \  +\frac{\gamma e^{-s}(x^\lambda - x^{1-\lambda})}{N} + \hdots,
\label{eq:1Nexpansion}
\end{align}
Recall from the main text that we write the elements of the maximal eigenvector as $(f_1, f_2, \hdots f_N)$ and
\begin{equation}
\begin{pmatrix}
f_i\\
f_{i+1}
\end{pmatrix} =
\begin{pmatrix}
\frac{\psi + 1 + x}{e^{-s} x^{1-\lambda}} & -x^{2\lambda - 1}\\
1 & 0
\end{pmatrix}
\begin{pmatrix}
f_{i+1}\\
f_{i+2}
\end{pmatrix} =
B \begin{pmatrix}
f_{i+1}\\
f_{i+2}
\end{pmatrix}.
\label{eq:Bdef2}
\end{equation}
Because of the translational symmetry of the network, the same transition matrix $B$ relates almost all pairs of neighboring $f_i$'s.
The heterogeneous link requires that we also introduce transfer matrices $A_1$ and $A_2$ given by
\begin{align}
\nonumber A_1 &=
\begin{pmatrix}
\frac{\psi + x + h_2}{e^{-s}h_1^{1-\lambda}h_2^\lambda} &
-\frac{x^{\lambda}}{h_1^{1-\lambda}h_2^\lambda}\\
1 & 0
\end{pmatrix}\\
A_2 &=
\begin{pmatrix}
\frac{\psi + h_1 + 1}{e^{-s}x^{1-\lambda}} &
-\frac{h_1^\lambda h_2^{1-\lambda}}{x^{1-\lambda}}\\
1 & 0
\end{pmatrix}.
\label{eq:A1A2}
\end{align}
Because the network is arranged in a ring, propagations around the full loop must map $(f_1, f_2)$ onto itself, such that the transfer matrices must satisfy the boundary condition
\begin{equation}
B^{N-2} A_2 A_1 \begin{pmatrix} f_1 \\ f_2 \end{pmatrix} = \begin{pmatrix} f_1 \\ f_2 \end{pmatrix},
\label{eq:boundarycondition}
\end{equation}
which requires that $B^{N-2} A_2 A_1$ posses a unit eigenvalue in the $N \to \infty$ limit.  We use this condition to determine the $1 / N$ expansion coefficient $\gamma$ as a function of $\lambda$ and $s$.
Since $\psi_{\omega, K}(\lambda,s) = \psi_{\omega, K}(1-\lambda, s)$ we focus on the case $x > 1$ and $\lambda < 1 / 2$ without loss of generality.
It is convenient to write $B$ in its eigenbasis after inserting the $1 / N$ expansion of Eq.~\eqref{eq:1Nexpansion} where only the larger of the two eigenvalues will survive the large $N$ limit,
\begin{equation}
\lim_{N \to \infty} B^{N-2} = \frac{e^{-\gamma}}{1 - x^{2 \lambda - 1}} \begin{pmatrix} 1 & -x^{2 \lambda - 1}\\ 1 & -x^{2 \lambda - 1}\end{pmatrix}
\label{BNm2}
\end{equation}

Since we are interested in the large $N$ behavior and there is only a single term of $A_1$ and $A_2$ in the product, we can comfortably neglect the $1/N$ term in the $A$ matrices. The condition that $B^{N-2}A_2 A_1$ has a unit eigenvalue requires
\begin{widetext}
\begin{equation}
\gamma = \ln \left[\frac{x^{2(1-\lambda)} + \left(-h_1 h_2 + e^{2s}(h_2 - 1)(h_1 - x)\right) + e^s(h_1 + h_2 - 1 - x)x^{1-\lambda}}{h_1^{1-\lambda}h_2^\lambda (x^{1-\lambda} - x^\lambda)}\right]
\label{eq:gamma}
\end{equation}
\end{widetext}
Note that $\gamma$ diverges when the numerator of the argument of the logarithm has a root, in which case $\psi_{\omega, K}(\lambda, s)$ departs significantly from the corresponding value in the translationally symmetric network.
The values of $\lambda$ and $s$ for which $\gamma$ first diverges provides the line of cusps given in Eq.~\eqref{eq:sstar} of the main text.
As shown in the main text, the value of $\psi_{\omega, K}(\lambda, s)$ everywhere follows.
When $\gamma$ does not diverge, the large $N$ behavior coincides with the translationally symmetric result.
Otherwise the behavior can be determined from the behavior of that translationally symmetric result along the curve $s^*(\lambda)$.


\begin{thebibliography}{40}%
\makeatletter
\providecommand \@ifxundefined [1]{%
 \@ifx{#1\undefined}
}%
\providecommand \@ifnum [1]{%
 \ifnum #1\expandafter \@firstoftwo
 \else \expandafter \@secondoftwo
 \fi
}%
\providecommand \@ifx [1]{%
 \ifx #1\expandafter \@firstoftwo
 \else \expandafter \@secondoftwo
 \fi
}%
\providecommand \natexlab [1]{#1}%
\providecommand \enquote  [1]{``#1''}%
\providecommand \bibnamefont  [1]{#1}%
\providecommand \bibfnamefont [1]{#1}%
\providecommand \citenamefont [1]{#1}%
\providecommand \href@noop [0]{\@secondoftwo}%
\providecommand \href [0]{\begingroup \@sanitize@url \@href}%
\providecommand \@href[1]{\@@startlink{#1}\@@href}%
\providecommand \@@href[1]{\endgroup#1\@@endlink}%
\providecommand \@sanitize@url [0]{\catcode `\\12\catcode `\$12\catcode
  `\&12\catcode `\#12\catcode `\^12\catcode `\_12\catcode `\%12\relax}%
\providecommand \@@startlink[1]{}%
\providecommand \@@endlink[0]{}%
\providecommand \url  [0]{\begingroup\@sanitize@url \@url }%
\providecommand \@url [1]{\endgroup\@href {#1}{\urlprefix }}%
\providecommand \urlprefix  [0]{URL }%
\providecommand \Eprint [0]{\href }%
\providecommand \doibase [0]{http://dx.doi.org/}%
\providecommand \selectlanguage [0]{\@gobble}%
\providecommand \bibinfo  [0]{\@secondoftwo}%
\providecommand \bibfield  [0]{\@secondoftwo}%
\providecommand \translation [1]{[#1]}%
\providecommand \BibitemOpen [0]{}%
\providecommand \bibitemStop [0]{}%
\providecommand \bibitemNoStop [0]{.\EOS\space}%
\providecommand \EOS [0]{\spacefactor3000\relax}%
\providecommand \BibitemShut  [1]{\csname bibitem#1\endcsname}%
\let\auto@bib@innerbib\@empty
\bibitem [{\citenamefont {Depken}\ \emph {et~al.}(2013)\citenamefont {Depken},
  \citenamefont {Parrondo},\ and\ \citenamefont {Grill}}]{Depken2013}%
  \BibitemOpen
  \bibfield  {author} {\bibinfo {author} {\bibfnamefont {M.}~\bibnamefont
  {Depken}}, \bibinfo {author} {\bibfnamefont {J.~M.}\ \bibnamefont
  {Parrondo}}, \ and\ \bibinfo {author} {\bibfnamefont {S.~W.}\ \bibnamefont
  {Grill}},\ }\href@noop {} {\bibfield  {journal} {\bibinfo  {journal} {Cell
  Reports}\ }\textbf {\bibinfo {volume} {5}},\ \bibinfo {pages} {521} (\bibinfo
  {year} {2013})}\BibitemShut {NoStop}%
\bibitem [{\citenamefont {Brun}\ \emph {et~al.}(2009)\citenamefont {Brun},
  \citenamefont {Rupp}, \citenamefont {Ward},\ and\ \citenamefont
  {N{\'e}d{\'e}lec}}]{Brun2009}%
  \BibitemOpen
  \bibfield  {author} {\bibinfo {author} {\bibfnamefont {L.}~\bibnamefont
  {Brun}}, \bibinfo {author} {\bibfnamefont {B.}~\bibnamefont {Rupp}}, \bibinfo
  {author} {\bibfnamefont {J.~J.}\ \bibnamefont {Ward}}, \ and\ \bibinfo
  {author} {\bibfnamefont {F.}~\bibnamefont {N{\'e}d{\'e}lec}},\ }\href@noop {}
  {\bibfield  {journal} {\bibinfo  {journal} {Proceedings of the National
  Academy of Sciences}\ }\textbf {\bibinfo {volume} {106}},\ \bibinfo {pages}
  {21173} (\bibinfo {year} {2009})}\BibitemShut {NoStop}%
\bibitem [{\citenamefont {Dogterom}\ and\ \citenamefont
  {Leibler}(1993)}]{Dogterom1993}%
  \BibitemOpen
  \bibfield  {author} {\bibinfo {author} {\bibfnamefont {M.}~\bibnamefont
  {Dogterom}}\ and\ \bibinfo {author} {\bibfnamefont {S.}~\bibnamefont
  {Leibler}},\ }\href@noop {} {\bibfield  {journal} {\bibinfo  {journal}
  {Physical Review Letters}\ }\textbf {\bibinfo {volume} {70}},\ \bibinfo
  {pages} {1347} (\bibinfo {year} {1993})}\BibitemShut {NoStop}%
\bibitem [{\citenamefont {Angelini}\ \emph {et~al.}(2011)\citenamefont
  {Angelini}, \citenamefont {Hannezo}, \citenamefont {Trepat}, \citenamefont
  {Marquez}, \citenamefont {Fredberg},\ and\ \citenamefont
  {Weitz}}]{Angelini2011}%
  \BibitemOpen
  \bibfield  {author} {\bibinfo {author} {\bibfnamefont {T.~E.}\ \bibnamefont
  {Angelini}}, \bibinfo {author} {\bibfnamefont {E.}~\bibnamefont {Hannezo}},
  \bibinfo {author} {\bibfnamefont {X.}~\bibnamefont {Trepat}}, \bibinfo
  {author} {\bibfnamefont {M.}~\bibnamefont {Marquez}}, \bibinfo {author}
  {\bibfnamefont {J.~J.}\ \bibnamefont {Fredberg}}, \ and\ \bibinfo {author}
  {\bibfnamefont {D.~A.}\ \bibnamefont {Weitz}},\ }\href@noop {} {\bibfield
  {journal} {\bibinfo  {journal} {Proceedings of the National Academy of
  Sciences}\ }\textbf {\bibinfo {volume} {108}},\ \bibinfo {pages} {4714}
  (\bibinfo {year} {2011})}\BibitemShut {NoStop}%
\bibitem [{\citenamefont {Balaban}\ \emph {et~al.}(2004)\citenamefont
  {Balaban}, \citenamefont {Merrin}, \citenamefont {Chait}, \citenamefont
  {Kowalik},\ and\ \citenamefont {Leibler}}]{Balaban2004}%
  \BibitemOpen
  \bibfield  {author} {\bibinfo {author} {\bibfnamefont {N.~Q.}\ \bibnamefont
  {Balaban}}, \bibinfo {author} {\bibfnamefont {J.}~\bibnamefont {Merrin}},
  \bibinfo {author} {\bibfnamefont {R.}~\bibnamefont {Chait}}, \bibinfo
  {author} {\bibfnamefont {L.}~\bibnamefont {Kowalik}}, \ and\ \bibinfo
  {author} {\bibfnamefont {S.}~\bibnamefont {Leibler}},\ }\href@noop {}
  {\bibfield  {journal} {\bibinfo  {journal} {Science}\ }\textbf {\bibinfo
  {volume} {305}},\ \bibinfo {pages} {1622} (\bibinfo {year}
  {2004})}\BibitemShut {NoStop}%
\bibitem [{\citenamefont {Touchette}(2009)}]{Touchette2009}%
  \BibitemOpen
  \bibfield  {author} {\bibinfo {author} {\bibfnamefont {H.}~\bibnamefont
  {Touchette}},\ }\href@noop {} {\bibfield  {journal} {\bibinfo  {journal}
  {Physics Reports}\ }\textbf {\bibinfo {volume} {478}},\ \bibinfo {pages} {1}
  (\bibinfo {year} {2009})}\BibitemShut {NoStop}%
\bibitem [{\citenamefont {Garrahan}\ \emph {et~al.}(2009)\citenamefont
  {Garrahan}, \citenamefont {Jack}, \citenamefont {Lecomte}, \citenamefont
  {Pitard}, \citenamefont {van Duijvendijk},\ and\ \citenamefont {van
  Wijland}}]{Garrahan2009}%
  \BibitemOpen
  \bibfield  {author} {\bibinfo {author} {\bibfnamefont {J.~P.}\ \bibnamefont
  {Garrahan}}, \bibinfo {author} {\bibfnamefont {R.~L.}\ \bibnamefont {Jack}},
  \bibinfo {author} {\bibfnamefont {V.}~\bibnamefont {Lecomte}}, \bibinfo
  {author} {\bibfnamefont {E.}~\bibnamefont {Pitard}}, \bibinfo {author}
  {\bibfnamefont {K.}~\bibnamefont {van Duijvendijk}}, \ and\ \bibinfo {author}
  {\bibfnamefont {F.}~\bibnamefont {van Wijland}},\ }\href@noop {} {\bibfield
  {journal} {\bibinfo  {journal} {Journal of Physics A: Mathematical and
  Theoretical}\ }\textbf {\bibinfo {volume} {42}},\ \bibinfo {pages} {075007}
  (\bibinfo {year} {2009})}\BibitemShut {NoStop}%
\bibitem [{\citenamefont {Espigares}\ \emph {et~al.}(2013)\citenamefont
  {Espigares}, \citenamefont {Garrido},\ and\ \citenamefont
  {Hurtado}}]{Espigares2013}%
  \BibitemOpen
  \bibfield  {author} {\bibinfo {author} {\bibfnamefont {C.~P.}\ \bibnamefont
  {Espigares}}, \bibinfo {author} {\bibfnamefont {P.~L.}\ \bibnamefont
  {Garrido}}, \ and\ \bibinfo {author} {\bibfnamefont {P.~I.}\ \bibnamefont
  {Hurtado}},\ }\href@noop {} {\bibfield  {journal} {\bibinfo  {journal}
  {Physical Review E}\ }\textbf {\bibinfo {volume} {87}},\ \bibinfo {pages}
  {032115} (\bibinfo {year} {2013})}\BibitemShut {NoStop}%
\bibitem [{\citenamefont {Bodineau}\ and\ \citenamefont
  {Derrida}(2004)}]{Bodineau2004}%
  \BibitemOpen
  \bibfield  {author} {\bibinfo {author} {\bibfnamefont {T.}~\bibnamefont
  {Bodineau}}\ and\ \bibinfo {author} {\bibfnamefont {B.}~\bibnamefont
  {Derrida}},\ }\href@noop {} {\bibfield  {journal} {\bibinfo  {journal}
  {Physical Review Letters}\ }\textbf {\bibinfo {volume} {92}},\ \bibinfo
  {pages} {180601} (\bibinfo {year} {2004})}\BibitemShut {NoStop}%
\bibitem [{\citenamefont {Bodineau}\ and\ \citenamefont
  {Toninelli}(2012)}]{Bodineau2012}%
  \BibitemOpen
  \bibfield  {author} {\bibinfo {author} {\bibfnamefont {T.}~\bibnamefont
  {Bodineau}}\ and\ \bibinfo {author} {\bibfnamefont {C.}~\bibnamefont
  {Toninelli}},\ }\href@noop {} {\bibfield  {journal} {\bibinfo  {journal}
  {Communications in Mathematical Physics}\ }\textbf {\bibinfo {volume}
  {311}},\ \bibinfo {pages} {357} (\bibinfo {year} {2012})}\BibitemShut
  {NoStop}%
\bibitem [{\citenamefont {Hedges}\ \emph {et~al.}(2009)\citenamefont {Hedges},
  \citenamefont {Jack}, \citenamefont {Garrahan},\ and\ \citenamefont
  {Chandler}}]{Hedges2009}%
  \BibitemOpen
  \bibfield  {author} {\bibinfo {author} {\bibfnamefont {L.~O.}\ \bibnamefont
  {Hedges}}, \bibinfo {author} {\bibfnamefont {R.~L.}\ \bibnamefont {Jack}},
  \bibinfo {author} {\bibfnamefont {J.~P.}\ \bibnamefont {Garrahan}}, \ and\
  \bibinfo {author} {\bibfnamefont {D.}~\bibnamefont {Chandler}},\ }\href@noop
  {} {\bibfield  {journal} {\bibinfo  {journal} {Science}\ }\textbf {\bibinfo
  {volume} {323}},\ \bibinfo {pages} {1309} (\bibinfo {year}
  {2009})}\BibitemShut {NoStop}%
\bibitem [{\citenamefont {Derrida}(2007)}]{Derrida2007}%
  \BibitemOpen
  \bibfield  {author} {\bibinfo {author} {\bibfnamefont {B.}~\bibnamefont
  {Derrida}},\ }\href@noop {} {\bibfield  {journal} {\bibinfo  {journal}
  {Journal of Statistical Mechanics: Theory and Experiment}\ }\textbf {\bibinfo
  {volume} {2007}},\ \bibinfo {pages} {P07023} (\bibinfo {year}
  {2007})}\BibitemShut {NoStop}%
\bibitem [{\citenamefont {Harris}\ \emph {et~al.}(2005)\citenamefont {Harris},
  \citenamefont {R{\'a}kos},\ and\ \citenamefont {Sch{\"u}tz}}]{Harris2005}%
  \BibitemOpen
  \bibfield  {author} {\bibinfo {author} {\bibfnamefont {R.}~\bibnamefont
  {Harris}}, \bibinfo {author} {\bibfnamefont {A.}~\bibnamefont {R{\'a}kos}}, \
  and\ \bibinfo {author} {\bibfnamefont {G.}~\bibnamefont {Sch{\"u}tz}},\
  }\href@noop {} {\bibfield  {journal} {\bibinfo  {journal} {Journal of
  Statistical Mechanics: Theory and Experiment}\ }\textbf {\bibinfo {volume}
  {2005}},\ \bibinfo {pages} {P08003} (\bibinfo {year} {2005})}\BibitemShut
  {NoStop}%
\bibitem [{\citenamefont {Vaikuntanathan}\ \emph {et~al.}(2014)\citenamefont
  {Vaikuntanathan}, \citenamefont {Gingrich},\ and\ \citenamefont
  {Geissler}}]{Vaikuntanathan2014}%
  \BibitemOpen
  \bibfield  {author} {\bibinfo {author} {\bibfnamefont {S.}~\bibnamefont
  {Vaikuntanathan}}, \bibinfo {author} {\bibfnamefont {T.~R.}\ \bibnamefont
  {Gingrich}}, \ and\ \bibinfo {author} {\bibfnamefont {P.~L.}\ \bibnamefont
  {Geissler}},\ }\href@noop {} {\bibfield  {journal} {\bibinfo  {journal}
  {Physical Review E}\ }\textbf {\bibinfo {volume} {89}},\ \bibinfo {pages}
  {062108} (\bibinfo {year} {2014})}\BibitemShut {NoStop}%
\bibitem [{\citenamefont {Garrahan}\ \emph {et~al.}(2007)\citenamefont
  {Garrahan}, \citenamefont {Jack}, \citenamefont {Lecomte}, \citenamefont
  {Pitard}, \citenamefont {van Duijvendijk},\ and\ \citenamefont {van
  Wijland}}]{Garrahan2007}%
  \BibitemOpen
  \bibfield  {author} {\bibinfo {author} {\bibfnamefont {J.~P.}\ \bibnamefont
  {Garrahan}}, \bibinfo {author} {\bibfnamefont {R.~L.}\ \bibnamefont {Jack}},
  \bibinfo {author} {\bibfnamefont {V.}~\bibnamefont {Lecomte}}, \bibinfo
  {author} {\bibfnamefont {E.}~\bibnamefont {Pitard}}, \bibinfo {author}
  {\bibfnamefont {K.}~\bibnamefont {van Duijvendijk}}, \ and\ \bibinfo {author}
  {\bibfnamefont {F.}~\bibnamefont {van Wijland}},\ }\href@noop {} {\bibfield
  {journal} {\bibinfo  {journal} {Physical Review Letters}\ }\textbf {\bibinfo
  {volume} {98}},\ \bibinfo {pages} {195702} (\bibinfo {year}
  {2007})}\BibitemShut {NoStop}%
\bibitem [{\citenamefont {Jack}\ and\ \citenamefont
  {Sollich}(2010)}]{Jack2010}%
  \BibitemOpen
  \bibfield  {author} {\bibinfo {author} {\bibfnamefont {R.~L.}\ \bibnamefont
  {Jack}}\ and\ \bibinfo {author} {\bibfnamefont {P.}~\bibnamefont {Sollich}},\
  }\href@noop {} {\bibfield  {journal} {\bibinfo  {journal} {Progress of
  Theoretical Physics}\ }\textbf {\bibinfo {volume} {184}},\ \bibinfo {pages} {304} (\bibinfo {year}
  {2010})}\BibitemShut {NoStop}%
\bibitem [{\citenamefont {Chetrite}\ and\ \citenamefont
  {Touchette}(2013)}]{Chetrite2013}%
  \BibitemOpen
  \bibfield  {author} {\bibinfo {author} {\bibfnamefont {R.}~\bibnamefont
  {Chetrite}}\ and\ \bibinfo {author} {\bibfnamefont {H.}~\bibnamefont
  {Touchette}},\ }\href@noop {} {\bibfield  {journal} {\bibinfo  {journal}
  {Physical Review Letters}\ }\textbf {\bibinfo {volume} {111}},\ \bibinfo
  {pages} {120601} (\bibinfo {year} {2013})}\BibitemShut {NoStop}%
\bibitem [{\citenamefont {Lewis}(2010)}]{Lewis2010}%
  \BibitemOpen
  \bibfield  {author} {\bibinfo {author} {\bibfnamefont {K.}~\bibnamefont
  {Lewis}},\ }\href@noop {} {\bibfield  {journal} {\bibinfo  {journal} {Annual
  Review of Microbiology}\ }\textbf {\bibinfo {volume} {64}},\ \bibinfo {pages}
  {357} (\bibinfo {year} {2010})}\BibitemShut {NoStop}%
\bibitem [{\citenamefont {Mitchison}\ and\ \citenamefont
  {Kirschner}(1984)}]{Mitchison1984}%
  \BibitemOpen
  \bibfield  {author} {\bibinfo {author} {\bibfnamefont {T.}~\bibnamefont
  {Mitchison}}\ and\ \bibinfo {author} {\bibfnamefont {M.}~\bibnamefont
  {Kirschner}},\ }\href@noop {} {\bibfield  {journal} {\bibinfo  {journal}
  {Nature}\ }\textbf {\bibinfo {volume} {312}},\ \bibinfo {pages} {237}
  (\bibinfo {year} {1984})}\BibitemShut {NoStop}%
\bibitem [{\citenamefont {Fujiwara}\ \emph {et~al.}(2002)\citenamefont
  {Fujiwara}, \citenamefont {Takahashi}, \citenamefont {Tadakuma},
  \citenamefont {Funatsu},\ and\ \citenamefont {Ishiwata}}]{Fujiwara2002}%
  \BibitemOpen
  \bibfield  {author} {\bibinfo {author} {\bibfnamefont {I.}~\bibnamefont
  {Fujiwara}}, \bibinfo {author} {\bibfnamefont {S.}~\bibnamefont {Takahashi}},
  \bibinfo {author} {\bibfnamefont {H.}~\bibnamefont {Tadakuma}}, \bibinfo
  {author} {\bibfnamefont {T.}~\bibnamefont {Funatsu}}, \ and\ \bibinfo
  {author} {\bibfnamefont {S.}~\bibnamefont {Ishiwata}},\ }\href@noop {}
  {\bibfield  {journal} {\bibinfo  {journal} {Nature Cell Biology}\ }\textbf
  {\bibinfo {volume} {4}},\ \bibinfo {pages} {666} (\bibinfo {year}
  {2002})}\BibitemShut {NoStop}%
\bibitem [{\citenamefont {Garner}\ \emph {et~al.}(2004)\citenamefont {Garner},
  \citenamefont {Campbell},\ and\ \citenamefont {Mullins}}]{Garner2004}%
  \BibitemOpen
  \bibfield  {author} {\bibinfo {author} {\bibfnamefont {E.~C.}\ \bibnamefont
  {Garner}}, \bibinfo {author} {\bibfnamefont {C.~S.}\ \bibnamefont
  {Campbell}}, \ and\ \bibinfo {author} {\bibfnamefont {R.~D.}\ \bibnamefont
  {Mullins}},\ }\href@noop {} {\bibfield  {journal} {\bibinfo  {journal}
  {Science}\ }\textbf {\bibinfo {volume} {306}},\ \bibinfo {pages} {1021}
  (\bibinfo {year} {2004})}\BibitemShut {NoStop}%
\bibitem [{\citenamefont {Popp}\ \emph {et~al.}(2010)\citenamefont {Popp},
  \citenamefont {Iwasa}, \citenamefont {Erickson}, \citenamefont {Narita},
  \citenamefont {Ma{\'e}da},\ and\ \citenamefont {Robinson}}]{Popp2010}%
  \BibitemOpen
  \bibfield  {author} {\bibinfo {author} {\bibfnamefont {D.}~\bibnamefont
  {Popp}}, \bibinfo {author} {\bibfnamefont {M.}~\bibnamefont {Iwasa}},
  \bibinfo {author} {\bibfnamefont {H.~P.}\ \bibnamefont {Erickson}}, \bibinfo
  {author} {\bibfnamefont {A.}~\bibnamefont {Narita}}, \bibinfo {author}
  {\bibfnamefont {Y.}~\bibnamefont {Ma{\'e}da}}, \ and\ \bibinfo {author}
  {\bibfnamefont {R.~C.}\ \bibnamefont {Robinson}},\ }\href@noop {} {\bibfield
  {journal} {\bibinfo  {journal} {Journal of Biological Chemistry}\ }\textbf
  {\bibinfo {volume} {285}},\ \bibinfo {pages} {11281} (\bibinfo {year}
  {2010})}\BibitemShut {NoStop}%
\bibitem [{\citenamefont {Dimitrov}\ \emph {et~al.}(2008)\citenamefont
  {Dimitrov}, \citenamefont {Quesnoit}, \citenamefont {Moutel}, \citenamefont
  {Cantaloube}, \citenamefont {Po{\"u}s},\ and\ \citenamefont
  {Perez}}]{Dimitrov2008}%
  \BibitemOpen
  \bibfield  {author} {\bibinfo {author} {\bibfnamefont {A.}~\bibnamefont
  {Dimitrov}}, \bibinfo {author} {\bibfnamefont {M.}~\bibnamefont {Quesnoit}},
  \bibinfo {author} {\bibfnamefont {S.}~\bibnamefont {Moutel}}, \bibinfo
  {author} {\bibfnamefont {I.}~\bibnamefont {Cantaloube}}, \bibinfo {author}
  {\bibfnamefont {C.}~\bibnamefont {Po{\"u}s}}, \ and\ \bibinfo {author}
  {\bibfnamefont {F.}~\bibnamefont {Perez}},\ }\href@noop {} {\bibfield
  {journal} {\bibinfo  {journal} {Science}\ }\textbf {\bibinfo {volume}
  {322}},\ \bibinfo {pages} {1353} (\bibinfo {year} {2008})}\BibitemShut
  {NoStop}%
\bibitem [{\citenamefont {Van~Kampen}(1992)}]{vankampen1992}%
  \BibitemOpen
  \bibfield  {author} {\bibinfo {author} {\bibfnamefont {N.~G.}\ \bibnamefont
  {Van~Kampen}},\ }\href@noop {} {\emph {\bibinfo {title} {Stochastic Processes
  in Physics and Chemistry}}},\ Vol.~\bibinfo {volume} {1}\ (\bibinfo
  {publisher} {Elsevier},\ \bibinfo {year} {1992})\BibitemShut {NoStop}%
\bibitem [{\citenamefont {Seifert}(2012)}]{Seifert2012}%
  \BibitemOpen
  \bibfield  {author} {\bibinfo {author} {\bibfnamefont {U.}~\bibnamefont
  {Seifert}},\ }\href@noop {} {\bibfield  {journal} {\bibinfo  {journal}
  {Reports on Progress in Physics}\ }\textbf {\bibinfo {volume} {75}},\
  \bibinfo {pages} {126001} (\bibinfo {year} {2012})}\BibitemShut {NoStop}%
\bibitem [{\citenamefont {Lebowitz}\ and\ \citenamefont
  {Spohn}(1999)}]{Lebowitz1999}%
  \BibitemOpen
  \bibfield  {author} {\bibinfo {author} {\bibfnamefont {J.~L.}\ \bibnamefont
  {Lebowitz}}\ and\ \bibinfo {author} {\bibfnamefont {H.}~\bibnamefont
  {Spohn}},\ }\href@noop {} {\bibfield  {journal} {\bibinfo  {journal} {Journal
  of Statistical Physics}\ }\textbf {\bibinfo {volume} {95}},\ \bibinfo {pages}
  {333} (\bibinfo {year} {1999})}\BibitemShut {NoStop}%
\bibitem [{\citenamefont {Touchette}\ and\ \citenamefont
  {Harris}(2013)}]{Touchette2013}%
  \BibitemOpen
  \bibfield  {author} {\bibinfo {author} {\bibfnamefont {H.}~\bibnamefont
  {Touchette}}\ and\ \bibinfo {author} {\bibfnamefont {R.~J.}\ \bibnamefont
  {Harris}},\ }\href@noop {} {\bibfield  {journal} {\bibinfo  {journal}
  {Nonequilibrium Statistical Physics of Small Systems: Fluctuation Relations
  and Beyond}\ ,\ \bibinfo {pages} {335}} (\bibinfo {year} {2013})}\BibitemShut
  {NoStop}%
\bibitem [{\citenamefont {Schnakenberg}(1976)}]{Schnakenberg1976}%
  \BibitemOpen
  \bibfield  {author} {\bibinfo {author} {\bibfnamefont {J.}~\bibnamefont
  {Schnakenberg}},\ }\href@noop {} {\bibfield  {journal} {\bibinfo  {journal}
  {Reviews of Modern Physics}\ }\textbf {\bibinfo {volume} {48}},\ \bibinfo
  {pages} {571} (\bibinfo {year} {1976})}\BibitemShut {NoStop}%
\bibitem [{\citenamefont {Gomez-Solano}\ \emph {et~al.}(2009)\citenamefont
  {Gomez-Solano}, \citenamefont {Petrosyan}, \citenamefont {Ciliberto},
  \citenamefont {Chetrite},\ and\ \citenamefont {Gaw{\c{e}}dzki}}]{Gomez2009}%
  \BibitemOpen
  \bibfield  {author} {\bibinfo {author} {\bibfnamefont {J.~R.}\ \bibnamefont
  {Gomez-Solano}}, \bibinfo {author} {\bibfnamefont {A.}~\bibnamefont
  {Petrosyan}}, \bibinfo {author} {\bibfnamefont {S.}~\bibnamefont
  {Ciliberto}}, \bibinfo {author} {\bibfnamefont {R.}~\bibnamefont {Chetrite}},
  \ and\ \bibinfo {author} {\bibfnamefont {K.}~\bibnamefont {Gaw{\c{e}}dzki}},\
  }\href@noop {} {\bibfield  {journal} {\bibinfo  {journal} {Physical Review
  Letters}\ }\textbf {\bibinfo {volume} {103}},\ \bibinfo {pages} {040601}
  (\bibinfo {year} {2009})}\BibitemShut {NoStop}%
\bibitem [{\citenamefont {Faggionato}\ \emph {et~al.}(2012)\citenamefont
  {Faggionato}, \citenamefont {Gabrielli} }]{Faggionato2012}%
  \BibitemOpen
  \bibfield  {author} {\bibinfo {author} {\bibfnamefont {A.}~\bibnamefont
  {Faggionato}} and \bibinfo {author} {\bibfnamefont {D.}~\bibnamefont
  {Gabrielli}} }\href@noop {} {\bibfield  {journal}
  {\bibinfo  {journal} {Annales de l'Institut Henri Poincar{\'e},
  Probabilit{\'e}s et Statistiques}\ }\textbf {\bibinfo {volume} {48}},\
  \bibinfo {pages} {212} (\bibinfo {year} {2012})}\BibitemShut {NoStop}%
\bibitem [{Note1()}]{Note1}%
  \BibitemOpen
  \bibinfo {note} {Localized eigenvectors of the tilted operator correspond to
  localized trajectories, which cannot produce entropy in the long-time limit.
  Partial derivatives with respect to $\lambda $ must then yield a zero entropy
  production.}\BibitemShut {Stop}%
\bibitem [{Note2()}]{Note2}%
  \BibitemOpen
  \bibinfo {note} {There is not a unique inverse, but the sum $x^{1-\lambda ^*}
  + x^{\lambda ^*}$ is the same for either choice of the inverse.}\BibitemShut
  {Stop}%
\bibitem [{Note3()}]{Note3}%
  \BibitemOpen
  \bibinfo {note} {Note that the cusp in the scaled cumulant generating
  function does not result from a simple eigenvalue crossing nor an avoided
  crossing as the delocalized vector ceases to be an eigenvector on the wrong
  side of the transition.}\BibitemShut {Stop}%
\bibitem [{\citenamefont {Bigger}(1944)}]{Bigger1944}%
  \BibitemOpen
  \bibfield  {author} {\bibinfo {author} {\bibfnamefont {J.}~\bibnamefont
  {Bigger}},\ }\href@noop {} {\bibfield  {journal} {\bibinfo  {journal} {The
  Lancet}\ }\textbf {\bibinfo {volume} {244}},\ \bibinfo {pages} {497}
  (\bibinfo {year} {1944})}\BibitemShut {NoStop}%
\bibitem [{\citenamefont {Lewis}(2000)}]{Lewis2000}%
  \BibitemOpen
  \bibfield  {author} {\bibinfo {author} {\bibfnamefont {K.}~\bibnamefont
  {Lewis}},\ }\href@noop {} {\bibfield  {journal} {\bibinfo  {journal}
  {Microbiology and Molecular Biology Reviews}\ }\textbf {\bibinfo {volume}
  {64}},\ \bibinfo {pages} {503} (\bibinfo {year} {2000})}\BibitemShut
  {NoStop}%
\bibitem [{\citenamefont {Keren}\ \emph {et~al.}(2004)\citenamefont {Keren},
  \citenamefont {Shah}, \citenamefont {Spoering}, \citenamefont {Kaldalu},\
  and\ \citenamefont {Lewis}}]{Keren2004}%
  \BibitemOpen
  \bibfield  {author} {\bibinfo {author} {\bibfnamefont {I.}~\bibnamefont
  {Keren}}, \bibinfo {author} {\bibfnamefont {D.}~\bibnamefont {Shah}},
  \bibinfo {author} {\bibfnamefont {A.}~\bibnamefont {Spoering}}, \bibinfo
  {author} {\bibfnamefont {N.}~\bibnamefont {Kaldalu}}, \ and\ \bibinfo
  {author} {\bibfnamefont {K.}~\bibnamefont {Lewis}},\ }\href@noop {}
  {\bibfield  {journal} {\bibinfo  {journal} {Journal of Bacteriology}\
  }\textbf {\bibinfo {volume} {186}},\ \bibinfo {pages} {8172} (\bibinfo {year}
  {2004})}\BibitemShut {NoStop}%
\bibitem [{\citenamefont {V{\'a}zquez-Laslop}\ \emph
  {et~al.}(2006)\citenamefont {V{\'a}zquez-Laslop}, \citenamefont {Lee},\ and\
  \citenamefont {Neyfakh}}]{Vazquez2006}%
  \BibitemOpen
  \bibfield  {author} {\bibinfo {author} {\bibfnamefont {N.}~\bibnamefont
  {V{\'a}zquez-Laslop}}, \bibinfo {author} {\bibfnamefont {H.}~\bibnamefont
  {Lee}}, \ and\ \bibinfo {author} {\bibfnamefont {A.~A.}\ \bibnamefont
  {Neyfakh}},\ }\href@noop {} {\bibfield  {journal} {\bibinfo  {journal}
  {Journal of Bacteriology}\ }\textbf {\bibinfo {volume} {188}},\ \bibinfo
  {pages} {3494} (\bibinfo {year} {2006})}\BibitemShut {NoStop}%
\bibitem [{\citenamefont {Bodineau}\ and\ \citenamefont
  {Derrida}(2005)}]{Bodineau2005}%
  \BibitemOpen
  \bibfield  {author} {\bibinfo {author} {\bibfnamefont {T.}~\bibnamefont
  {Bodineau}}\ and\ \bibinfo {author} {\bibfnamefont {B.}~\bibnamefont
  {Derrida}},\ }\href@noop {} {\bibfield  {journal} {\bibinfo  {journal}
  {Physical Review E}\ }\textbf {\bibinfo {volume} {72}},\ \bibinfo {pages}
  {066110} (\bibinfo {year} {2005})}\BibitemShut {NoStop}%
\bibitem [{\citenamefont {Bertini}\ \emph {et~al.}(2005)\citenamefont
  {Bertini}, \citenamefont {De~Sole}, \citenamefont {Gabrielli}, \citenamefont
  {Jona-Lasinio},\ and\ \citenamefont {Landim}}]{Bertini2005}%
  \BibitemOpen
  \bibfield  {author} {\bibinfo {author} {\bibfnamefont {L.}~\bibnamefont
  {Bertini}}, \bibinfo {author} {\bibfnamefont {A.}~\bibnamefont {De~Sole}},
  \bibinfo {author} {\bibfnamefont {D.}~\bibnamefont {Gabrielli}}, \bibinfo
  {author} {\bibfnamefont {G.}~\bibnamefont {Jona-Lasinio}}, \ and\ \bibinfo
  {author} {\bibfnamefont {C.}~\bibnamefont {Landim}},\ }\href@noop {}
  {\bibfield  {journal} {\bibinfo  {journal} {Physical Review Letters}\
  }\textbf {\bibinfo {volume} {94}},\ \bibinfo {pages} {030601} (\bibinfo
  {year} {2005})}\BibitemShut {NoStop}%
\bibitem [{\citenamefont {Bertini}\ \emph {et~al.}(2006)\citenamefont
  {Bertini}, \citenamefont {De~Sole}, \citenamefont {Gabrielli}, \citenamefont
  {Jona-Lasinio},\ and\ \citenamefont {Landim}}]{Bertini2006}%
  \BibitemOpen
  \bibfield  {author} {\bibinfo {author} {\bibfnamefont {L.}~\bibnamefont
  {Bertini}}, \bibinfo {author} {\bibfnamefont {A.}~\bibnamefont {De~Sole}},
  \bibinfo {author} {\bibfnamefont {D.}~\bibnamefont {Gabrielli}}, \bibinfo
  {author} {\bibfnamefont {G.}~\bibnamefont {Jona-Lasinio}}, \ and\ \bibinfo
  {author} {\bibfnamefont {C.}~\bibnamefont {Landim}},\ }\href@noop {}
  {\bibfield  {journal} {\bibinfo  {journal} {Journal of Statistical Physics}\
  }\textbf {\bibinfo {volume} {123}},\ \bibinfo {pages} {237} (\bibinfo {year}
  {2006})}\BibitemShut {NoStop}%
\end{thebibliography}
%

\end{document}